\documentclass[aps,twocolumn,showpacs,preprintnumbers,nofootinbib,prd,
superscriptaddress,groupedaddress,10pt]{revtex4-2}
\usepackage[utf8]{inputenc}
\usepackage{graphicx,amssymb,amsmath,amsthm,amsfonts,epsfig,times,natbib}
\usepackage[usenames,dvipsnames]{color}
\usepackage{epstopdf}
\usepackage{aas_macros}
\usepackage{tensor}
\usepackage{mathtools}
\usepackage{mathrsfs}
\usepackage{amsbsy}
\usepackage{bm,url}
\usepackage[linktocpage]{hyperref}
\usepackage[scaled]{beramono}
\usepackage[T1]{fontenc}
\usepackage{mathrsfs}
\usepackage{ulem}
\usepackage{soul}
\usepackage[usenames]{color}

\newcommand{\be}{\begin{equation}}
\newcommand{\ee}{\end{equation}}
\newcommand{\nn}{\nonumber}

\definecolor{oxfordblue}{rgb}{0.0, 0.13, 0.28}
\definecolor{burgundy}{rgb}{0.5, 0.0, 0.13}
\definecolor{darkolivegreen}{rgb}{0.33, 0.42, 0.18}
\definecolor{darkblue}{rgb}{0,0,0.5}
\definecolor{richcarmine}{rgb}{0.84, 0.0, 0.25}
\definecolor{darkblue}{rgb}{0,0,0.5}
\definecolor{venetianred}{rgb}{0.78, 0.03, 0.08}
\definecolor{skobeloff}{rgb}{0.0, 0.48, 0.45}
\hypersetup{colorlinks=true, citecolor=darkblue, linkcolor=darkblue,
urlcolor = darkblue}

\def\nn{\nonumber}

\newcommand{\ben}{\begin{enumerate}}
\newcommand{\een}{\end{enumerate}}

\def\be{\begin{equation}}
\def\ee{\end{equation}}
\def\nn{\nonumber}
\newcommand{\beq}{\begin{eqnarray}}
\newcommand{\eeq}{\end{eqnarray}}
\newcommand{\ba}{\begin{array}}
\newcommand{\ea}{\end{array}}

\begin{document}

\title{Extreme mass-ratio inspirals around a spinning horizonless compact object}

\author{
Elisa Maggio$^{1}$,
Maarten van de Meent$^{2}$,
Paolo Pani$^{1}$
}
\affiliation{${^1}$  Dipartimento di Fisica, ``Sapienza'' Universit\`a di Roma \& Sezione INFN Roma1, Piazzale
Aldo Moro 5, 00185, Roma, Italy}
\affiliation{${^2}$ Max Planck Institute for Gravitational Physics (Albert Einstein Institute) Am M\"{u}hlenberg 1, 14476 Potsdam, Germany}

\begin{abstract}
Extreme mass-ratio inspirals~(EMRIs) detectable by the Laser Interferometer Space Antenna are unique probes of the nature of supermassive compact objects.
We compute the gravitational-wave signal emitted by a stellar-mass compact object in a circular equatorial orbit around a Kerr-like horizonless supermassive object defined by an effective radius and a reflectivity coefficient. The Teukolsky equations are solved consistently with suitable (frequency-dependent) boundary conditions, and the modified energy and angular-momentum fluxes are used to evolve the orbital parameters adiabatically. The gravitational fluxes have resonances corresponding to the low-frequency quasinormal modes of the central object, which can contribute significantly to the gravitational-wave phase. Overall, the absence of a classical event horizon in the central object can affect the gravitational-wave signal dramatically, with deviations even larger than those previously estimated by a model-independent analysis of the tidal heating. We estimate that EMRIs could potentially place the most stringent constraint on the reflectivity of supermassive compact objects at the remarkable level of ${\cal O}(10^{-6})\%$ and would allow one to constrain various models which are not ruled out by the ergoregion instability. In particular, an EMRI detection could allow one to rule out (or provide evidence for) signatures of quantum black-hole horizons with Boltzmann reflectivity. Our results provide motivation for performing rigorous parameter estimation to assess the detectability of these effects.
\end{abstract}

\maketitle
%

\section{Introduction}

The defining feature of a classical black hole~(BH) is its function as a perfect absorber since its event horizon is a one-way, null hypersurface.
Thus, any evidence of some partial \emph{reflectivity} near a dark compact object would 
indicate a departure from the classical BH picture, while an upper bound on the reflectivity could help quantify the ``BH-ness'' of a 
dark compact source~\cite{Cardoso:2019rvt}.

Gravitational-wave~(GW) astronomy naturally provides the ideal setting to constrain the reflectivity of compact sources.
In fact, one might argue that imperfect GW absorption should be the rule rather than the exception since event horizons are very special and all known forms of matter interact very weakly with GWs, even in extreme conditions~\cite{1971ApJ...165..165E,1966ApJ...145..544H,PhysRevD.96.084033,PhysRevD.99.123030}.
On the contrary, owing to their horizon, BHs are dissipative systems that behave like a 
Newtonian viscous fluid~\cite{MembraneParadigm,Damour_viscous,Poisson:2009di,Cardoso:2012zn}. 
A spinning Kerr BH absorbs radiation of frequency $\omega>m\Omega_H$ (where $m$ is the azimuthal number of the 
wave and $\Omega_H$ is the BH angular velocity) but amplifies radiation of smaller frequency, due to 
superradiance (see~\cite{Brito:2015oca} for a review).

In the last few years, several studies have explored the possibility of constraining the reflectivity of compact GW sources (see Refs.~\cite{Cardoso:2019rvt,Maggio:2021ans} for some reviews), mostly modeling the postmerger ``echo'' signal from exotic compact objects~(ECOs)~\cite{Cardoso:2016rao,Cardoso:2016oxy,Abedi:2016hgu,Mark:2017dnq,Cardoso:2017cqb,Nakano:2017fvh,Wang:2018gin,Testa:2018bzd,Maggio:2019zyv,Abedi:2020ujo,Maggio:2020jml}
or deriving projected bounds on the so-called tidal heating~--~namely, the dissipative\footnote{This dissipative effect should not be confused with the effect of the tidal deformability of a compact object, encoded in its tidal Love numbers, which also affects the conservative part of the dynamics. Tidal heating is associated with the energy and angular-momentum absorption by the compact object and results in an increase of the mass and angular momentum of the latter, unless superradiance~\cite{Brito:2015oca} occurs, in which case the energy and angular-momentum fluxes have opposite signs (as discussed in detail below). See Ref.~\cite{PoissonWill} for an introduction on tidal effects in a binary system and Ref.~\cite{Cardoso:2019rvt} for their different roles in tests of ECOs.} backreaction on the orbital motion from the tides that were raised during the coalescence~\cite{Hartle:1973zz,Hughes:2001jr,PoissonWill}.
In a comparable-mass binary, tidal heating enters the GW phase at high post-Newtonian order~\cite{Alvi:2001mx} and is therefore hard to measure~\cite{Maselli:2017cmm}.
On the other hand, tidal heating in extreme mass-ratio inspirals~(EMRIs) can produce thousands of radians of accumulated 
orbital phase~\cite{Hughes:2001jr,Bernuzzi:2012ku,Taracchini:2013wfa,Harms:2014dqa,Datta:2019epe} while in the sensitivity band of the future 
space-based Laser~Interferometer~Space~Antenna~(LISA)~\cite{LISA:2017pwj}.
Recently, this effect was studied to develop a test that can place very stringent and model-independent constraints on the reflectivity of supermassive objects~\cite{Datta:2019epe,Datta:2020rvo},
which adds to other unparalleled EMRI-based tests of fundamental physics, such as no-hair theorem tests 
based on measurements of the multipolar structure of the central object~\cite{Ryan:1995wh,Barack:2006pq,Babak:2017tow, Datta:2019epe, Bianchi:2020bxa,Bah:2021jno}, constraints on extra degrees of freedom arising in modified gravity~\cite{Sopuerta:2009iy,Yunes:2011aa,Pani:2011xj,Cardoso:2018zhm,Maselli:2020zgv}, and null-hypothesis tests based on the absence of tidal Love numbers~\cite{Pani:2019cyc}. Together, these tests 
suggest that EMRIs will be unique probes of the nature of supermassive objects (for recent reviews on these and 
other tests, see Refs.~\cite{Gair:2012nm,Cardoso:2019rvt,Baibhav:2019rsa,Barausse:2020rsu}).

The phenomenological approach of Ref.~\cite{Datta:2019epe} was to study a standard BH EMRI dynamics and to parametrize a certain amount of reflectivity at the object surface in terms of a constant reflectivity coefficient $|{\cal R}|^2$, assuming that a fraction $(1-|{\cal R}|^2)$ of the radiation is absorbed. Clearly, the BH limit is recovered as ${\cal R}\to0$, whereas $|{\cal R}|^2=1$ corresponds to a perfectly reflective object. According to the analysis in~\cite{Datta:2019epe}, EMRIs could provide an unparalleled constraint at the level of $|{\cal R}|^2\lesssim10^{-4}$, much more stringent than current and future echo searches~\cite{Testa:2018bzd,Maggio:2019zyv,Cardoso:2019rvt,Abedi:2020ujo}.

The main goal of our paper is to improve the analysis of Ref.~\cite{Datta:2019epe} by studying a consistent model of a compact horizonless object, defined by a certain compactness and (possibly frequency-dependent) reflectivity coefficient, which in turn modify the boundary conditions for radiation at the surface. In our model, tidal heating and partial reflection are not imposed by hand, but rather
arise automatically from the boundary conditions. The latter also generically affect the dynamics as well as the quasinormal modes~(QNMs) of the central object~\cite{Pani:2009ss,Maggio:2017ivp,Maggio:2018ivz,Maggio:2020jml}. The QNM spectrum typically contains low-frequency modes arising from long-lived, quasibound states, which might be resonantly excited during the inspiral~\cite{Pani:2010em,Macedo:2013jja,Cardoso:2019nis,Fransen:2020prl}. 
The role of these resonances in the EMRI dynamics was studied in Ref.~\cite{Cardoso:2019nis} for a perfectly reflecting, nonspinning, quasi-Schwarzschild horizonless object, for which an analytical treatment of the problem is possible (see also Ref.~\cite{Fransen:2020prl} for a more recent study). Our framework allows one to extend the analysis of Ref.~\cite{Cardoso:2019nis} to the case of a generic (and possibly frequency-dependent) reflectivity coefficient and generic spin. As we shall show, at variance with the case studied in Ref.~\cite{Cardoso:2019nis}, in more generic situations the presence of resonances can provide an important contribution to the EMRI dynamics.
We shall also show that, by taking a consistent model into account, the already very stringent potential bounds derived by Ref.~\cite{Datta:2019epe} can be further improved by some orders of magnitude. Finally, we show that EMRI detections have the potential to rule out (or provide observational hints of) models of quantum-gravity BH horizons featuring a (frequency-dependent) Boltzmann reflectivity~\cite{Oshita:2019sat}.

The rest of this paper is organized as follows. In Sec.~\ref{sec:setup} we present our analytical and numerical framework, which relies on solving the EMRI dynamics around an ECO to leading order in an adiabatic expansion. We present our results in Sec.~\ref{sec:results} and conclude in Sec.~\ref{sec:discussion}.
Through this work, we use $G=c=1$ units.

\section{Setup} \label{sec:setup}

\subsection{A model for a Kerr-like horizonless object}

We analyze a spinning compact horizonless object whose exterior spacetime is described by the Kerr metric~\cite{Maggio:2017ivp,Abedi:2016hgu,Barausse:2018vdb}.
It is worth remarking that, even within general relativity, the vacuum region outside a spinning object is not necessarily described by the Kerr geometry due to the absence of Birkhoff's theorem beyond spherical symmetry. However, in the BH limit, any deviation from the multipolar structure of a Kerr BH dies off sufficiently fast~\cite{Raposo:2018xkf} within general relativity or 
in modified theories of gravity whose effects are confined near the radius of the compact object\footnote{\label{footnote}
For EMRIs, assuming that the central object is described by the Kerr metric is also justified for gravity theories with higher-curvature/high-energy corrections to general relativity~\cite{Berti:2015itd}. In that case, the corrections to the metric are suppressed by powers of $l_P/r_0 \ll 1$, where $r_0$ is the radius of the central object and $l_P$ is the Planck length or the length scale of new physics~\cite{Maselli:2020zgv,Piovano:2020zin}.
}.
Explicit examples of this ``hair-conditioner theorem''~\cite{Raposo:2018xkf} within general relativity are given in 
Refs.~\cite{Pani:2015tga,Uchikata:2015yma,Uchikata:2016qku,Yagi:2015hda,Yagi:2015upa,Posada-Aguirre:2016qpz}.

Therefore, within our framework a spinning horizonless object with compactness close to the BH one can be approximated by the Kerr metric in the exterior spacetime and the properties of the interior can be modeled in terms of a reflectivity coefficient.

In Boyer-Lindquist coordinates, the line element outside the compact object reads
\begin{eqnarray}
ds^2&&=-\left(1-\frac{2Mr}{\Sigma}\right)dt^2+\frac{\Sigma}{\Delta}dr^2-\frac{
4Mr}{\Sigma}a\sin^2\theta d\phi dt   \nn \\
&+&{\Sigma}d\theta^2+
\left[(r^2+a^2)\sin^2\theta +\frac{2Mr}{\Sigma}a^2\sin^4\theta
\right]d\phi^2\,,\label{Kerr}
\end{eqnarray}
where $\Sigma=r^2+a^2\cos^2\theta$ and $\Delta=r^2+a^2-2M r$, with $M$ and $J\equiv aM$ the total mass and angular momentum of the object, respectively.
We shall consider a horizonless compact object whose radius is located at
\begin{equation}
 r_0 = r_+(1+\epsilon)\,,  \label{epsilon-def}
\end{equation}
where $r_+=M+\sqrt{M^2-a^2}$ is the location of the would-be horizon. Let us notice that the parameter $\epsilon$ is related to the compactness of the object namely,
$M/r_0 \approx M/r_+ (1-\epsilon)$ when $\epsilon\ll1$. Motivated by models of microscopic corrections at the horizon scale, we shall focus mostly on the case in which $\epsilon\ll1$.
For example,  if $r_0\sim r_+ +l_P$ (where $l_P$ is the Planck length, as suggested by some
quantum-gravity inspired models~\cite{Guo:2017jmi}), then
$\epsilon\sim 10^{-44}$ for a compact object with $M \sim 10^6 \,
M_{\odot}$ and spin $a/M=0.9$.

The properties of the interior are parametrized in terms of a complex and frequency-dependent reflectivity coefficient $\mathcal{R}$. The $\mathcal{R}=0$ case describes a totally absorbing compact object (which reduces to the standard BH case when $\epsilon\to0$), whereas the $|\mathcal{R}|^2=1$ case describes a perfectly reflecting compact object. Intermediate values of $\mathcal{R}$ describe partially absorbing compact objects due to viscosity or dissipation within the object~\cite{Mark:2017dnq,Maggio:2017ivp,Maggio:2018ivz,Maggio:2019zyv,Oshita:2019sat,Maggio:2020jml}.

\subsection{Linear perturbations from a point particle} \label{sec:perturbation}

Let us consider the case of a pointlike source orbiting around a central object (either a Kerr BH or a Kerr-like ECO) in a circular equatorial orbit.
In line with the previous discussion, we assume that in the exterior of the object general relativity is valid, at least approximately (this does not prevent beyond-general-relativity corrections in the object interior and at the horizon scale, which can be parametrized by the reflectivity coefficient). Therefore, gravitational perturbations in the exterior can be described as in the Kerr BH case.
We analyze the gravitational perturbation in the Newman-Penrose formalism.\footnote{Very recently, using the Sasaki-Nakamura perturbations, a similar formalism was used to study a point particle plunging onto a  spinning compact horizonless object in the context of developing accurate echo waveforms~\cite{Xin:2021zir} (see also Ref.~\cite{Sago:2020avw}).} The Weyl scalar $\Psi_4$ can be expanded as 
\begin{equation}
\Psi_4 = \hat\rho^4 \sum_{\ell,m} \int_{-\infty}^{\infty} d\omega R_{\ell m \omega}(r) {}_{-2}S_{\ell m \omega}(\theta) e^{i (m \phi - \omega t)} \,,
\end{equation}
where $\hat\rho = (r-ia \cos \theta)^{-1}$ and the sum runs over $\ell \geq 2$ and $-\ell \leq m \leq \ell$. The radial wave function $R_{\ell m \omega}(r)$ and the spin-weighted spheroidal harmonics ${}_{-2}S_{\ell m \omega}(\theta) e^{i m \phi}$ obey to the Teukolsky master equations~\cite{Teukolsky:1972my,Teukolsky:1973ha,Teukolsky:1974yv}
\begin{eqnarray}
&&\Delta^{2} \frac{d}{dr}\left(\frac{1}{\Delta} \frac{dR_{\ell m \omega}}{dr}\right) - V(r) R_{\ell m \omega} = \mathcal{T}_{\ell m \omega} \,,\label{wave_eq} \\
&&\bigg[ \frac{1}{\sin \theta} \frac{d}{d \theta} \left( \sin \theta \frac{d}{d \theta}\right) + a^2 \omega^2 \cos^2 \theta - \left( \frac{m-2 \cos \theta}{\sin \theta}\right)^2 \nonumber\\
&+& 4 a \omega \cos \theta -4 + {}_{-2}A_{\ell m \omega}\bigg]{}_{-2}S_{\ell m \omega}=0 \,, \label{eq_angular}
\end{eqnarray}
where the effective potential reads
\begin{equation}
V(r) = - \frac{K^{2}+4 i (r-M) K}{\Delta}+8 i \omega r +\lambda \,,
\end{equation}
where $K=(r^2+a^2)\omega-am$, and the separation
constants $\lambda$ and ${}_{-2}A_{\ell m \omega}$ are related by $\lambda \equiv  {}_{-2}A_{\ell m}-2am\omega+a^2\omega^2-2$.
The polar part of the spin-weighted spheroidal harmonics is normalized such that
\begin{equation}
\int_{-1}^{1} \left|{}_{-2}S_{\ell m \omega}(\cos \theta)\right|^2 d\cos\theta =1 \,.
\end{equation}
The source term $\mathcal{T}_{\ell m \omega}$ is constructed by projecting the stress-energy tensor $T^{\alpha \beta}$ of a pointlike source with respect to the Newman-Penrose tetrad, where~\cite{Fujita:2004rb}
\begin{equation}
T^{\alpha \beta} = \mu \frac{u^{\alpha} u^{\beta}}{\Sigma \sin \theta u^t} \delta\left(r-r(t)\right) \delta\left(\theta-\theta(t)\right) \delta\left(\phi-\phi(t)\right) \,,
\end{equation}
where $\mu$ is the mass of the small orbiting body, $u^{\alpha} = dz^{\alpha}/d\tau$, $z^{\alpha} = \left( t, r(t), \theta(t), \phi(t)\right)$ is the geodesic trajectory, and $\tau$ is the particle's proper time. We define the mass ratio of the system as $q=\mu/M$. In the case of circular equatorial orbits, $\theta(t) = \pi/2$ and, for corotating orbits, the orbital radius is related to the orbital angular frequency by
\begin{equation}
\Omega = \sqrt{M}/(a \sqrt{M} + r^{3/2}) \,. \label{orbfreq}
\end{equation}

Equation~\eqref{wave_eq} can be solved through the standard Green's function method using the solutions of the homogeneous Teukolsky equation.

\subsubsection{BH case}
Let us first review the standard BH case. Owing to the presence of a horizon, the two independent homogeneous solutions have the following asymptotic behavior:
\begin{equation} \label{asymptoticsin}
R^{\rm in}_{\ell m \omega} \sim 
\begin{cases}
 \displaystyle 
B^{\rm trans}_{\ell m \omega} \Delta^2 e^{-i k r_*} & \text{ as } r_* \to - \infty \,, \\ 
 \displaystyle  
 r^3 B^{\rm ref}_{\ell m \omega}  e^{i \omega r_*}  +  r^{-1} B^{\rm inc}_{\ell m \omega} e^{- i \omega r_*} & \text{ as } r_* \to 
+ \infty \,,
\end{cases} \\
\end{equation}
\begin{equation} \label{asymptoticsup}
R^{\rm up}_{\ell m \omega} \sim 
\begin{cases}
 \displaystyle 
 C^{\rm up}_{\ell m \omega} e^{i k r_*}  +  \Delta^2 C^{\rm ref}_{\ell m \omega} e^{-i k r_*} & \text{ as } r_* \to - \infty \,, \\ 
 \displaystyle  
 r^3 C^{\rm trans}_{\ell m \omega} e^{i \omega r_*} & \text{ as } r_* \to + \infty \,, \\
\end{cases}
\end{equation}
where $k = \omega - m \Omega_H$, $\Omega_H = a/(2Mr_+)$ is the angular velocity at the horizon of the Kerr BH and the tortoise coordinate is defined such that $dr_*/dr = (r^2 + a^2)/\Delta$. 
The inhomogeneous solution of the Teukolsky equation~\eqref{wave_eq} is constructed as~\cite{Fujita:2004rb}
\begin{eqnarray}
\nonumber R_{\ell m \omega} &=& \frac{1}{W_{\ell m \omega}} \left\{ R^{\rm up}_{\ell m \omega}(r) \int_{r_+}^r dr' \frac{\mathcal{T}_{\ell m \omega}(r') R^{\rm in}_{\ell m \omega}(r')}{\Delta^2(r')} \right. \\
&+& \left. R^{\rm in}_{\ell m \omega}(r) \int_{r}^{\infty} dr' \frac{\mathcal{T}_{\ell m \omega}(r') R^{\rm up}_{\ell m \omega}(r')}{\Delta^2(r')} \right\} \,, \label{inhomsol}
\end{eqnarray}
where $W_{\ell m \omega}$ is the Wronskian given by
\begin{eqnarray}
W_{\ell m \omega} &=& \Delta^{-1} \left( R^{\rm in}_{\ell m \omega} \frac{dR^{\rm up}_{\ell m \omega}}{dr} - R^{\rm up}_{\ell m \omega} \frac{dR^{\rm in}_{\ell m \omega}}{dr} \right) \nonumber \\
&=& 2 i \omega C^{\rm trans}_{\ell m \omega} B^{\rm inc}_{\ell m \omega} \,. 
\end{eqnarray}
The inhomogeneous solution in Eq.~\eqref{inhomsol} has the following asymptotic behavior:
\begin{equation}
R_{\ell m \omega} \sim 
\begin{cases}
 \displaystyle 
 Z_{\ell m \omega}^{H} \Delta^2 e^{-i k r_*} & \text{ as } r_* \to - \infty \,, \\ 
 \displaystyle  
 Z_{\ell m \omega}^{\infty} r^3 e^{i \omega r_*} & \text{ as } r_* \to + \infty \,, \\
\end{cases}
\end{equation}
where
\begin{eqnarray}
Z_{\ell m \omega}^{H} &=& C^{H}_{\ell m \omega} \int_{r_+}^{\infty} dr' \frac{\mathcal{T}_{\ell m \omega}(r') R^{\rm up}_{\ell m \omega}(r')}{\Delta^2(r')} \,, \\
Z_{\ell m \omega}^{\infty} &=& C^{\infty}_{\ell m \omega}  \int_{r_+}^{\infty} dr' \frac{\mathcal{T}_{\ell m \omega}(r') R^{\rm in}_{\ell m \omega}(r')}{\Delta^2(r')} \,,
\end{eqnarray}
and
\begin{equation}
C^{H}_{\ell m \omega} = \frac{B^{\rm trans}_{\ell m \omega}}{2 i \omega C^{\rm trans}_{\ell m \omega} B^{\rm inc}_{\ell m \omega}} \,, \quad C^{\infty}_{\ell m \omega} = \frac{1}{2 i \omega B^{\rm inc}_{\ell m \omega}} \,.
\end{equation}

The amplitudes $Z_{\ell m \omega}^{H}$ and $Z_{\ell m \omega}^{\infty}$ determine the gravitational energy fluxes emitted at infinity and through the horizon~\cite{Teukolsky:1974yv,Hughes:1999bq}:
\begin{eqnarray}
\dot{E}^{\infty} = \sum_{\ell m} \frac{|Z^{\infty}_{\ell m \omega}|^2}{4 \pi (m \Omega)^2} \,, \label{Einf}\\
\dot{E}^{H} = \sum_{\ell m} \frac{\alpha_{\ell m} |Z^{H}_{\ell m \omega}|^2}{4 \pi (m \Omega)^2} \label{Eh}\,,
\end{eqnarray}
where
\begin{equation}
\alpha_{\ell m} = \frac{256 (2Mr_+)^5 k (k^2 + 4 \varpi^2) (k^2 + 16 \varpi^2) (m \Omega)^3}{|c_{\ell m}|^2} \,,
\end{equation}
with $\varpi = \sqrt{M^2-a^2}/(4Mr_+)$ and

\begin{eqnarray}
|c_{\ell m}|^2 &=& [(\lambda+2)^2 + 4 m a (m\Omega) - 4a^2 (m\Omega)^2 ] \nonumber \\
& \times & [\lambda^2 + 36 m a (m\Omega) - 36 a^2 (m\Omega)^2] \nonumber \\
& + & (2 \lambda +3) [96 a^2 (m\Omega)^2 - 48 m a (m\Omega)] \nonumber \\
& + & 144 (m\Omega)^2 (M^2-a^2) \,.
\end{eqnarray}
For circular equatorial orbits, the angular-momentum fluxes are related to the energy fluxes at infinity and at the horizon by $\dot{J}^{\infty, H} = \dot{E}^{\infty, H}/\Omega$.

In the BH case, the total energy flux emitted by a point particle in a  circular equatorial orbit with orbital angular frequency $\Omega$ is
\begin{equation}
 \dot E(\Omega) = \dot E^{\infty}(\Omega) +\dot E^{H}(\Omega)\,, \label{EtotBH}
\end{equation}
where $\dot E^{\infty}(\Omega)$ and $\dot E^{H}(\Omega)$ are as  defined in Eqs.~\eqref{Einf} and~\eqref{Eh}, respectively.

\subsubsection{Horizonless case} \label{sec:horinzoless}
As discussed above, we assume that, at least in the exterior of the central object, general relativity is a valid approximation. Therefore, for $r>r_0$ the perturbations equations are the same as in the Kerr BH case, and possible corrections can be incorporated into the boundary conditions for the gravitational radiation at the effective radius.

The physical interpretation of the inner boundary condition is more evident by adopting the Detweiler function~\cite{1977RSPSA.352..381D}
\begin{equation}
X_{\ell m \omega} = \frac{\left(r^2+a^2\right)^{1/2}}{\Delta} \left[\alpha(r) \
R_{\ell m \omega}+\beta(r) \Delta^{-1} \frac{dR_{\ell m \omega}}{dr}\right]\,,\label{DetweilerX}
\end{equation}
where $\alpha(r)$ and $\beta(r)$ are certain radial functions~\cite{1977RSPSA.352..381D,Maggio:2018ivz}.
Indeed, since any signal is totally absorbed, in the BH case the physical solution is a purely ingoing wave near the horizon, 
\begin{equation}
X_{\ell m \omega} \sim e^{-i k r_*} \quad \text{ as } r_* \to - \infty \,.
\end{equation}

In the case of an ECO, the solution near the surface ($r\sim r_0$) is more involved and also depends on the value of $\epsilon$~\cite{Maggio:2020jml}.
If we assume that $\epsilon \ll 1$, the effective potential in the
Detweiler equation is constant near the surface, $V\approx-k^2$~\cite{1977RSPSA.352..381D,Maggio:2018ivz}, so the perturbation is a superposition of ingoing and outgoing waves at the ECO radius,
\begin{equation}
 X_{\ell m \omega} \sim A_{\rm in} e^{-i k r_*} + A_{\rm out}  e^{ik r_*} \quad \text{ as } r_* \to r_*^0 \,,
\end{equation}
where $r_*^0 \equiv r_*(r_0)$. One can therefore define the surface reflectivity of the ECO as~\cite{Maggio:2018ivz}
\begin{equation}
\mathcal{R(\omega)} = \frac{A_{\rm out}}{A_{\rm in}} e^{2 i k r_*^0} \,.
\end{equation}
Perfectly reflecting objects have $|\mathcal{R}(\omega)|^2=1$, i.e., ${\cal R}(\omega)=e^{i\psi(\omega)}$ for an arbitrary (real) frequency-dependent phase $\psi$. Two notable examples of perfectly reflecting boundary conditions are
\begin{equation} 
 \left\{\begin{array}{ll}
    X_{\ell m \omega}(r_0)=0 &\quad \rm Dirichlet, \ \mathcal{R}=-1 \,, \\
    dX_{\ell m \omega}(r_0)/dr_*=0 &\quad \rm Neumann, \ \mathcal{R}=1 \,, \\
   \end{array}\right.
\end{equation}
corresponding to $\psi=\pi$ and $\psi=0$, respectively. In general, a partially absorbing compact object is described by
\begin{equation}
\left. \frac{dX_{\ell m \omega}/dr_*}{X_{\ell m \omega}} \right|_{r_0}= - i k \frac{1-\mathcal{R(\omega)}}{1+\mathcal{R(\omega)}} \,, \label{BC}
\end{equation}
which reduces to the BH boundary condition when ${\cal R}=0$.

In the ECO case, the solutions of the homogeneous Teukolsky equation are such that the ``up'' modes have the same asymptotics as Eq.~\eqref{asymptoticsup}, whereas the ``in'' modes have the following asymptotics:
\begin{equation} \label{RinECO}
R^{\rm in}_{\ell m \omega} \sim 
\begin{cases}
 \displaystyle 
B'^{\rm trans}_{\ell m \omega} \Delta^2 e^{-i k r_*}
+ C'^{\rm up}_{\ell m \omega} e^{i k r_*} 
& \text{ as } r_* \to r_*^0 \,, \\
 \displaystyle  
 r^3 B'^{\rm ref}_{\ell m \omega} e^{i \omega r_*}
 +  r^{-1} B'^{\rm inc}_{\ell m \omega} e^{- i \omega r_*}
& \text{ as } r_* \to + \infty \,,
\end{cases} \\
\end{equation}
where
\begin{eqnarray}
B'^{\rm trans}_{\ell m \omega} &=& B^{\rm trans}_{\ell m \omega}+c_1 C^{\rm ref}_{\ell m \omega} \,, \\
C'^{\rm up}_{\ell m \omega} &=& c_1 C^{\rm up}_{\ell m \omega} \,, \\
B'^{\rm ref}_{\ell m \omega} &=& B^{\rm ref}_{\ell m \omega}+c_1 C^{\rm trans}_{\ell m \omega} \,, \\
B'^{\rm inc}_{\ell m \omega}  &=& B^{\rm inc}_{\ell m \omega} \,,
\end{eqnarray}
and the coefficient $c_1$ is determined by imposing the boundary condition in Eq.~\eqref{BC} with
\begin{equation}
R_{\ell m \omega} = R^{\rm in}_{\ell m \omega} + c_1 R^{\rm up}_{\ell m \omega} \,.
\end{equation}
The inhomogeneous solution of the Teukolsky function is derived as in Eq.~\eqref{inhomsol}, with $R^{\rm in}_{\ell m \omega}$ as in Eq.~\eqref{RinECO} and $R^{\rm up}_{\ell m \omega}$ as in Eq.~\eqref{asymptoticsup}, and it has the following asymptotic behavior:
\begin{equation} \label{inhomECO}
R_{\ell m \omega} \sim 
\begin{cases}
 \displaystyle 
 Z_{\ell m \omega}^{H^+} \Delta^2 e^{-i k r_*} + Z_{\ell m \omega}^{H^-}  e^{i k r_*} & \text{ as } r_* \to r_*^0 \,, \\ 
 \displaystyle  
 Z_{\ell m \omega}^{\infty} r^3 e^{i \omega r_*} & \text{ as } r_* \to + \infty \,, \\
\end{cases} 
\end{equation}
where
\begin{equation}
Z_{\ell m \omega}^{H^+} = Z_{\ell m \omega}^{H} \,, \quad
Z_{\ell m \omega}^{H^-} = \frac{{C'^{\rm up}_{\ell m \omega}}}{B'^{\rm trans}_{\ell m \omega}} Z_{\ell m \omega}^{H} \,.
\end{equation}
To determine the energy emitted by the particle in the ECO case, we note that -- by assumption -- the gravitational perturbations in the neighborhood of the particle are exactly those of a Kerr background, albeit with unusual boundary conditions. We can therefore determine the emitted energy by appealing to the energy balance law in the Kerr background. The energy flux to infinity is formally given by the same formula as the BH case, Eq.~\eqref{Einf}. The energy flux to the ECO side is determined by analytically extending $R_{\ell m \omega}$ to the horizon of the Kerr background and measuring the flux there. Thus, the internal energy flux on the ECO side $\dot{E}^{\rm int}$ is given by
\begin{equation}
\dot{E}^{\rm int} = \dot{E}^{H^{+}} - \dot{E}^{H^{-}} \label{Eradius} \,,
\end{equation}
where $\dot{E}^{H^{+}}$ and $\dot{E}^{H^{-}}$ are the energy fluxes across the future and past horizon, respectively. The flux across the future horizon is as in the BH case given by Eq.~\eqref{Eh}, while the energy flux coming in across the past horizon is~\cite{Teukolsky:1974yv}\footnote{Note that while~\cite{Teukolsky:1974yv} calculates only the flux across the future horizon, the flux across the past horizon can be obtained trivially by switching the roles of the ingoing and outgoing principal null vectors, which has the effect of reversing the roles of $\psi_0$ and $\psi_4$.}
\begin{equation}
\dot{E}^{H^{-}} =\sum_{\ell m} \frac{\omega}{4 \pi k (2{M}r_+)^3 (k^2 + 4 \varpi^2)} |Z^{H^{-}}_{\ell m \omega}|^2 \,.
\end{equation}
In the case of $\mathcal{R}=0$, Eq.~\eqref{Eradius} reduces to $\dot{E}^{\rm int} = \dot{E}^{H^{+}}$.
When $|\mathcal{R}(\omega)|^2=1$, the outgoing flux is equal to the ingoing flux at the ECO radius and $\dot{E}^{\rm int} = 0$, as expected from fully reflecting boundary conditions. 

In the ECO case, the total energy flux emitted by a point particle in a  circular equatorial orbit is
\begin{equation}
 \dot E(\Omega) = \dot E^{\infty}(\Omega) +\dot{E}^{\rm int}(\Omega)\,, \label{EtotECO}
\end{equation}
where $\dot E^{\infty}(\Omega)$ and $\dot{E}^{\rm int}(\Omega)$ are as defined in Eqs.~\eqref{Einf} and~\eqref{Eradius}, respectively.

\subsection{Adiabatic evolution and waveform} \label{sec:adiabatic}

In an EMRI, the radiation-reaction timescale is much longer than the orbital period so --~at the first order in the mass ratio~-- the orbital parameters can be evolved using an adiabatic expansion~\cite{Hinderer:2008dm}. For a particle in a circular equatorial and corotating orbit,  
the evolution of the orbital angular frequency $\Omega$ and the orbital phase $\phi$ are governed by
\begin{eqnarray}
 \dot \Omega &=& -\left(\frac{d E_b}{d \Omega}\right)^{-1} \dot E(\Omega)\,, \label{eq:adiab1}\\
 \dot \phi&=& \Omega \label{eq:adiab2}\,,
\end{eqnarray}
where $E_b$ is the binding energy of the system, 
\begin{equation}
E_b = \mu \frac{1-2 v^2 + \chi v^3}{\sqrt{1-3v^2+2\chi v^3}} \,,
\end{equation}
where $\chi=a/M$, $v \equiv \sqrt{M/r}$, $r$ is the orbital radius which is related to the orbital angular frequency via Eq.~\eqref{orbfreq}, and $\dot E(\Omega)$ is the total energy flux defined in Eqs.~\eqref{EtotBH} and~\eqref{EtotECO} in the BH and in the ECO case, respectively.

Equations~\eqref{eq:adiab1} and~\eqref{eq:adiab2} can be solved with some initial conditions $\Omega(t=0)=\Omega_0$ and (without loss of generality) $\phi(t=0)=0$. The GW phase of the dominant mode is related to the orbital phase by $\phi_{\rm GW} = 2 \phi$. The GW dephasing accumulated up to a certain time between the BH case and ECO case is computed as~\cite{Datta:2019epe}
\begin{equation}
\Delta \phi(t) = \phi_{\rm GW}^{\rm BH}(t) - \phi_{\rm GW}^{\rm ECO}(t) \,.
\end{equation}

The emitted waveform is computed from the Weyl scalar at infinity and reads~\cite{Hughes:2001jr,Piovano:2020zin}
\begin{eqnarray}
 h_+- i h_\times =& \displaystyle-\frac{2}{\sqrt{2 \pi}} \frac{\mu}{D}  \sum_{\ell m} \frac{Z_{\ell m \omega}^{\infty}(t)}{\left[m \Omega(t)\right]^2} e^{i m\left( \Omega(t) r_*^D-\phi(t)\right)} \nonumber \\
&\times ~_{-2}S_{\ell m \omega}(\vartheta,t) e^{i m \varphi} \,, \label{waveform}
\end{eqnarray}
where $D$ is the source luminosity distance from the detector, $r_*^D \equiv r_*(D)$, 
and $(\vartheta,\varphi)$ identify the direction, in Boyer-Lindquist coordinates, 
of the detector in a reference frame centered at the source. 
Since the initial phase is degenerate with the azimuthal direction, we simply rescale the initial phase as $\varphi \equiv \phi(t=0)$.

Note that, regardless of its reflectivity, an ultracompact object can efficiently trap radiation within its photon 
sphere~\cite{Cardoso:2014sna,Cardoso:2016rao,Cardoso:2016oxy}. If radiation is trapped for enough time, it can contribute to the energy balance used to evolve the orbit adiabatically, thus mimicking the effect of a horizon even in the absence of dissipation within the object. However, this energy trapping at the photon sphere is not effective for a particle in circular orbit if~\cite{Maselli:2017cmm,Datta:2019epe}
\begin{equation}
 \epsilon \gg \exp\left(- \frac{5\sqrt{1-\chi^2}}{64q\left(1+\sqrt{1-\chi^2}\right)}\right)\,.
\end{equation}
Owing to the mass-ratio dependence, this condition is always satisfied in the EMRI limit ($q \lesssim 10^{-5}$) for any realistic value of $\epsilon$ and $\chi$.
Therefore, for an EMRI the only way to absorb radiation near the central ECO is by dissipating in its interior, i.e., when $|{\cal R}|^2<1$.

Spinning horizonless Kerr-like objects are affected by the so-called ergoregion instability~\cite{1978CMaPh..63..243F,10.1093/mnras/282.2.580} when spinning sufficiently fast~\cite{Cardoso:2007az,Cardoso:2008kj,Pani:2010jz,Maggio:2017ivp,Maggio:2018ivz}. In this case, the central object would spin down, reaching a stable configuration. The instability timescale can be shorter than the orbital period and would affect the dynamics of the point particle. Since unstable solutions should not form in the first place and, in any case, do not live long enough to form an EMRI, we shall focus our analysis on stable Kerr-like horizonless objects only. Stability is reached by assuming partially absorbing compact objects ($\mathcal{R}<1$)~\cite{Maggio:2018ivz} or specific models for the frequency-dependent reflectivity $\mathcal{R}(\omega)$~\cite{Oshita:2019sat}, so the net absorption of the relevant frequencies is higher than the superradiant amplification that leads to the ergoregion instability~\cite{Maggio:2017ivp,Maggio:2018ivz}.

Finally, note that horizonless compact objects contain low-frequency modes in their spectrum, which are associated with long-lived quasibound states efficiently confined within the object photon sphere~\cite{Cardoso:2014sna,Maggio:2017ivp,Maggio:2018ivz,Maggio:2020jml}. Unlike the BH case, these low-frequency modes can be excited during a quasicircular inspiral when the orbital frequency equals the QNM frequency, leading to resonances in the fluxes~\cite{Pani:2010em,Macedo:2013jja,Cardoso:2019nis,Fransen:2020prl}. As discussed in Sec.~\ref{sec:results}, these resonances can be very narrow and require very high resolution in order to resolve them.

\subsection{Overlap}\label{sec:overlap}

Although the dephasing $\Delta\phi$ between two different waveforms [$h_1(t)$ and $h_2(t)$] is a useful and quick measure to estimate the impact of different effects, a somewhat
more reliable and robust measure for assessing the measurability of any deviation from a standard reference signal is given by the overlap:
\begin{equation}\label{overlap}
\mathit{O}(h_1|h_2) = \frac{\left\langle h_1|h_2\right\rangle}{\sqrt{\left\langle h_1|h_1\right\rangle \left\langle h_2|h_2\right\rangle}}\,,
\end{equation}
where the inner product $\left\langle h_1|h_2\right\rangle$ is defined by
\begin{equation}
\left\langle h_1|h_2\right\rangle = 4\Re\,\int_{0}^{\infty} \frac{\tilde{h}_1 \tilde{h}^*_2}{S_n(f)} df\,,
\end{equation}
and $S_n(f)$ is the GW detector noise power spectral density, and the quantities with tildes and the star stand for the Fourier transform and complex conjugation, respectively. For the power spectral density, we adopted the LISA curve of Ref.~\cite{Cornish:2018dyw} adding the contribution of the confusion noise from the unresolved Galactic binaries for a one-year mission lifetime.
Since the waveforms 
are defined up to an arbitrary time and phase shift, it is also necessary to maximize the overlap in Eq.~\eqref{overlap} over 
these quantities. In practice, this can be done by computing~\cite{Allen:2005fk} 
\begin{equation}\label{overlap2}
\mathcal{O}(h_1|h_2) = \frac{4}{\sqrt{\left\langle h_1|h_1\right\rangle \left\langle h_2|h_2\right\rangle}}\max_{t_0} 
\left|\mathcal{F}^{-1}\left[\frac{\tilde{h}_1 \tilde{h}^*_2}{S_n(f)}\right](t_0)\right|\,,
\end{equation}
where $\mathcal{F}^{-1}[g(f)](t) =\int_{-\infty}^{+\infty} g(f) e^{-2\pi i f t}df$ is the inverse Fourier 
transform. The overlap is defined such that $\mathcal{O}=1$ indicates perfect agreement between two waveforms. 
It is also customary to define the mismatch $\mathcal{M}\equiv 1-{\cal O}$.

\subsection{Numerical procedure} \label{sec:numproc}

We have studied the dynamics of a point particle in a circular equatorial orbit around a Kerr-like ECO by adapting the frequency-domain Teukolsky code originally developed in Refs.~\cite{vandeMeent:2014raa, vandeMeent:2015lxa, vandeMeent:2016pee, vandeMeent:2017bcc}. In particular, the solutions to the homogeneous Teukolsky equation are calculated via the numerical Mano-Suzuki-Takasugi method~\cite{Mano:1996vt, Mano:1996gn, Fujita:2004rb,Fujita:2009us}. Use of this method gives full analytical control over the boundary conditions, making it perfectly suited for our purpose. We have modified the (frequency-dependent) boundary conditions at $r=r_0$ in terms of ${\cal R}$ and $\epsilon$, as discussed in Sec.~\ref{sec:perturbation}, computed the energy and angular-momentum fluxes at infinity and through the object's surface, and finally evolved the quasicircular orbit adiabatically by integrating Eqs.~\eqref{eq:adiab1} and~\eqref{eq:adiab2}. 

Our algorithm is as follows:
\begin{itemize}
\item[1.] Choose the intrinsic parameters of the binary--namely, the central mass $M$, the mass ratio $q\ll1$, the primary spin $\chi$, the reflectivity $\mathcal{R}(\omega)$, and the compactness of the central object and the initial orbital radius.
\item[2.] For a given $\ell=m$ mode, produce the data for a bound orbit with orbital radius $r$ and compute the energy fluxes in the cases of a central BH and a central ECO, respectively.
\item[3.] Loop on the orbital radii with an equally spaced (radial) grid starting with the ISCO radius to $r=10M$.  
\item[4.] Find the local maxima and minima in the energy fluxes at infinity for a central ECO. If present, these extrema bracket resonances in the flux, which should be resolved by increasing the grid resolution. Note that the initial equally spaced grid in the orbital radii needs to be dense enough to find local maxima and minima. For this reason, we set the initial discretization in the orbital radii at $0.003M$.
\item[5.] Refine the grid on the orbital radii around local maxima and minima through bisection until a target accuracy is reached. The refinement of the grid stops either when the difference between two subsequent orbital radii is $<10^{-5}M$ or when the difference in the energy fluxes of two subsequent points is $<10^{-5}q^2$.
\item[6.] For a given $\ell$ and each $m = \ell-1, ..., 1$ loop on the orbital radii with an equally spaced grid from the ISCO radius. 
The loop on the orbital radii stops when the total energy flux in the case of a central BH [defined in Eq.~\eqref{EtotBH}] in the given $\ell, m$ mode is $10^{-6}$ times smaller than the total energy flux in the dominant mode with $\ell=m$.
\item[7.] For a given $\ell$ and each $m = \ell-1, ..., 1$ repeat steps 4 and 5.
\item[8.] For the harmonic index $\ell=2, ..., \ell_{\rm max}=12$ repeat the steps 2 to 7.
\item[9.] For each $\ell, m$ mode, interpolate the total energy flux as a function of the orbital angular frequency.
\item[10.] Sum over the modes and integrate Eqs.~\eqref{eq:adiab1} and~\eqref{eq:adiab2} to compute the orbital phase in both the BH and  ECO cases. The initial condition on the orbital angular frequency is $\Omega_0 = \Omega(r=10M)$ and the integration stops at the inspiral-plunge transition frequency~\cite{Ori:2000zn} $\Omega(t_{\rm max}) = \Omega(r=r_{\rm ISCO} + 4q^{2/3})$.
\end{itemize}

The gravitational waveform is computed via Eq.~\eqref{waveform}, where for the modes with negative $m$ we make use of the following symmetries:
\begin{eqnarray}
Z^{\infty}_{\ell -m \omega} &=& (-1)^{\ell} \left(Z^{\infty}_{\ell m \omega}\right)^* \,, \\
~_{-2}S_{\ell -m \omega}(\vartheta) &=& (-1)^{\ell} ~_{-2}S_{\ell m \omega}(\pi-\vartheta) \,.
\end{eqnarray}
For each $\ell, m$ mode the asymptotic amplitudes at infinity and the spin-weighted spheroidal harmonics are interpolated functions of the time-dependent orbital angular frequency. The waveform is constructed by summing over the modes with $\ell \leq 4$ and $-\ell \leq m \leq \ell$. In the cases of small reflectivity ($|\mathcal{R}|^2\leq 10^{-6}$) the waveform is constructed by summing over the $\ell, m$ modes until $\ell = 5$ 
since one needs higher accuracy to keep the truncation errors smaller than the ECO corrections. We checked to see that the mismatch between the BH and ECO waveforms does not change quantitatively by including modes with higher $\ell $ in the waveforms.

We tested our code by reproducing standard results for the Kerr BH case~\cite{Hughes:2001jr,Bernuzzi:2012ku,Taracchini:2013wfa,Harms:2014dqa}. Furthermore, we reproduced the results of Ref.~\cite{Datta:2019epe} for a Kerr-like ECO using the same assumptions; i.e., we considered the Kerr BH case and artificially impose the requirement that only a fraction $(1-|{\cal R}|^2)$ of the radiation is absorbed at the surface.

The fractional truncation error of the code is estimated in the dephasing as 
$\Delta^{\rm tr} = 1- \Delta \phi_{\ell_{\rm max}+1}(t_f) / \Delta \phi_{\ell_{\rm max}}(t_f)$, where the energy fluxes are truncated at $\ell_{\rm max}=12$ and $t_f$ is the time at which the orbital radius is $r=r_{\rm ISCO} + 4 q^{2/3}$. For a reference compact object with $\chi=0.9$, $|\mathcal{R}|^2=0.9$, $\epsilon=10^{-10}$, and $q=3 \times 10^{-5}$, we find that $\Delta^{\rm tr}=2 \times 10^{-5}$.

\section{Results}\label{sec:results}
In this section, we present our main results. In Secs.~\ref{sec:energyfluxes}-~\ref{sec:resoverlap} we consider an agnostic model with generic values of $\epsilon$ and ${\cal R}$. In Sec.~\ref{sec:Boltzmann}, we specialize to the Boltzmann reflectivity model of Ref.~\cite{Oshita:2019sat}.

\subsection{Energy fluxes and resonances}\label{sec:energyfluxes}

%
\begin{figure}[ht]
\centering
\includegraphics[width=0.49\textwidth]{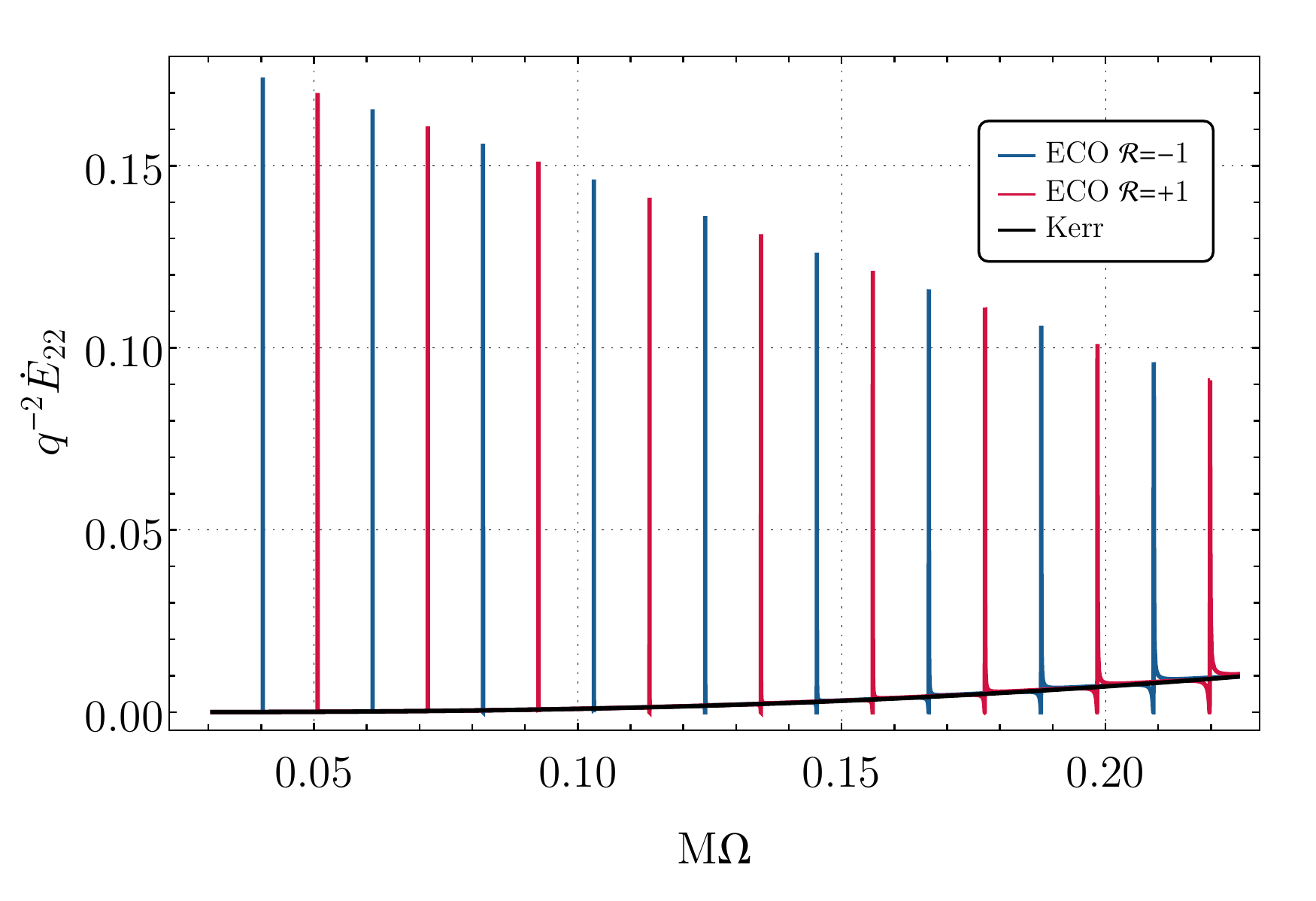}
\caption{Total energy flux of the $\ell=m=2$ mode as a function of the orbital angular frequency for a point particle in quasicircular equatorial orbit from $r=10M$ to $r=r_{\rm ISCO}$. We compare the case of a central Kerr BH with spin $\chi=0.9$ to the case of a central ECO with a perfectly reflecting surface ($|\mathcal{R}|^2=1$), $\chi=0.9$, and $\epsilon=10^{-10}$. In the latter case, the flux is resonantly excited when the orbital frequency matches the low-frequency QNMs of the ECO.} 
\label{fig:flux}
\end{figure}

Let us start by discussing the modified energy flux in the case of a spinning horizonless compact object.
As a representative example, Fig.~\ref{fig:flux} shows the
$\ell=m=2$ component of the total energy flux as a function of the orbital frequency for $\epsilon=10^{-10}$, $\chi=0.9$, and two choices (Dirichlet and Neumann) of perfectly reflecting boundary conditions. 
As expected, the flux is resonantly excited when the frequency matches the low-frequency QNMs of the central ECO, i.e., $\Omega = \omega_R/m$, where $\omega_R$ and $\omega_I$ are the real and the imaginary part of the QNMs, respectively. This is a striking difference with respect to the BH case, since the Kerr QNMs have higher frequencies and cannot be resonantly excited by quasicircular inspirals. In the small-$\epsilon$ limit, the Dirichlet and Neumann modes are described by~\cite{Maggio:2018ivz}
\begin{eqnarray}
\omega_R&\sim&
- \frac{\pi (p+1)}{2|r_*^0|}+m\Omega_H \,,
\label{wRana_grav} \\
 \omega_I &\sim&
-\frac{\beta_{\ell}}{|r_*^0|}\left(\frac{2 M r_+}{r_+-r_-}\right)\left[
\omega_R(r_+-r_-)\right]^{2\ell+1}(\omega_R-m\Omega_H)\,, \nn\\\label{wIana_grav}
\end{eqnarray}
where $r_\pm=M \pm\sqrt{M^2-a^2}$, $\sqrt{\beta_{\ell}}=\frac{(\ell+2)! (\ell-2)!}{(2\ell)! (2\ell+1)!!}$, and $p$ is an odd (even) integer for Neumann (Dirichlet) modes. 
As shown in Fig.~\ref{fig:flux}, for fixed $\chi$ and $\epsilon$ the modes are equispaced with $\Delta \omega_R=\pi/|r_*^0|$, whereas consecutive Dirichlet and Neumann mode frequencies are separated by half this width. The difference between consecutive resonances scales as $\Delta \omega_R \sim |\log \epsilon|^{-1}$, and therefore the resonances are denser in the $\epsilon \to 0$ limit. 

Interestingly, the resonances appear at the same frequencies in all the individual fluxes: $\dot E^\infty$, $\dot E^{H^+}$, and $\dot E^{H^-}$. This is due to the fact that the QNMs are associated with the poles of the Wronskian appearing in each solution of Teukolsky's equation. However, when $|{\cal R}|^2=1$ the fluxes $\dot E^{H^+}$ and $\dot E^{H^-}$ are exactly equal to each other since in this case $\dot{E}^{\rm int}=0$. Therefore, for the perfectly reflecting case resonances appear only in the flux at infinity.

Equation~\eqref{wIana_grav} shows that $\omega_I\ll\omega_R$, which implies that the resonances are typically very narrow and hard to resolve~\cite{Pani:2010em,Cardoso:2019nis,Fransen:2020prl}. 
The energy flux across a single resonance is very well fitted by a forced harmonic oscillator model~\cite{Pons:2001xs}
\begin{equation}
\frac{\dot{E}^{\rm ECO}}{\dot{E}^{\rm BH}} = \frac{[(1-b) (m\Omega)^2 - \omega_R^2 - \omega_I^2]^2+(2 m\Omega \omega_I)^2}{\left[(m\Omega)^2 - \omega_R^2 - \omega_I^2\right]^2 + (2 m\Omega \omega_I)^2} \,,
\end{equation}
where $\dot{E}$ is the total energy flux as computed in Eqs.~\eqref{EtotBH} and~\eqref{EtotECO}, respectively, for the BH and ECO cases, $b = 1-(\Omega_{\rm max}/\Omega_{\rm min})^2$, and $\Omega_{\rm max}$ and $\Omega_{\rm min}$ are the orbital angular frequencies of the maximum and the minimum of each resonance.
The width of each resonance in the orbital frequency scales as $\delta \Omega \sim \omega_I$~\cite{Cardoso:2019nis}, where $\omega_I \sim \omega_R^{2\ell+2}$ from Eq.~\eqref{wIana_grav}. It follows that the width of the resonances increases with the orbital angular frequency as shown in Fig.\ref{fig:flux}. In the nonspinning and perfectly reflecting case, we recover the results of Ref.~\cite{Cardoso:2019nis}--namely, that low-frequency resonances do not contribute significantly to the GW phase (see Appendix~\ref{app:dephasing_spin0}). 
However, as discussed below, for highly spinning compact objects, the ISCO frequency occurs at higher frequencies with respect to the nonspinning case and higher-frequency resonances can be efficiently excited, contributing a significant dephasing with respect to the BH case. 

\begin{figure}[ht]
\centering
\includegraphics[width=0.49\textwidth]{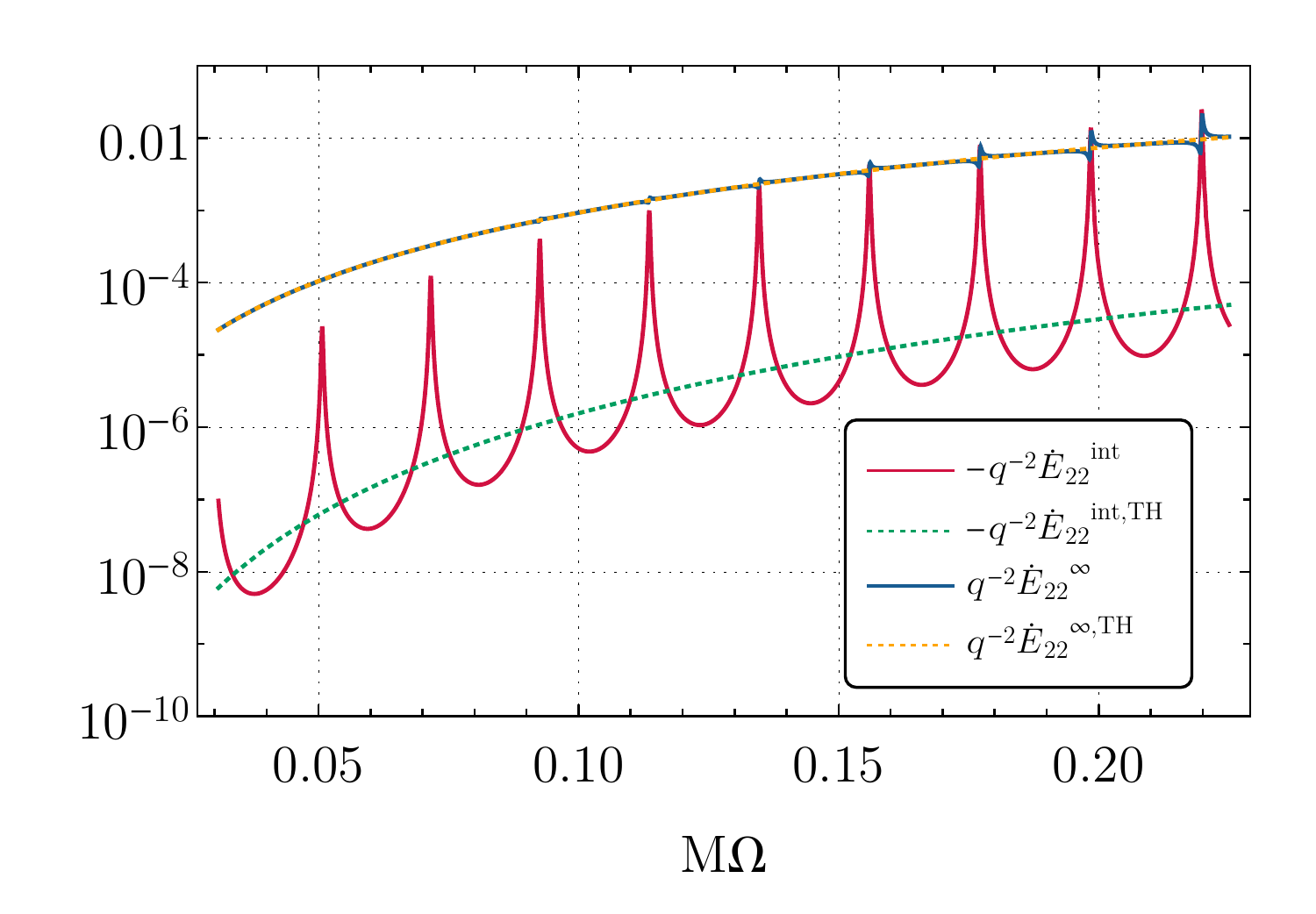}
\caption{Energy fluxes that are emitted at the radius and at infinity by a point particle around a central ECO with $\chi=0.9$, $\epsilon=10^{-10}$, and $\mathcal{R}=\sqrt{0.9}$ for the $\ell=m=2$ mode. The fluxes are compared to those of Ref.~\cite{Datta:2019epe} in which the effect of the ECO was accounted for by simply removing a fraction $(|{\cal R}|^2)$ of the tidal heating~(TH) from a standard Kerr EMRI flux.} 
\label{fig:flux2}
\end{figure}

The system shown in Fig.~\ref{fig:flux} is purely indicative since for this choice of the parameters the central ECO is unstable and would tend to spin down on short timescales~\cite{Maggio:2017ivp,Maggio:2018ivz}. This is also shown by the fact that $\omega_I$ in Eq.~\eqref{wIana_grav} is positive ($\omega_R<m\Omega_H$), as expected due to the ergoregion instability.
Stable solutions require either smaller values of the spin or partial absorption~\cite{Maggio:2017ivp,Maggio:2018ivz}. In all these cases the resonances are less evident, as shown in Fig.~\ref{fig:flux2}, where we considered a model with $|{\cal R}|^2=0.9$, a value that guarantees stability for $\chi=0.9$~\cite{Maggio:2018ivz}. 

Several comments are in order.
First, also for a smaller reflectivity we observe resonances in $\dot E^\infty$ as in the perfectly reflecting case of Fig.~\ref{fig:flux}: in this case, they are less peaked but, as shown below, could still have a sufficiently large width to be efficiently excited. Second, the same resonant frequencies appear also in $\dot{E}^{\rm int}$. This is due to the fact that for $|{\cal R}|<1$ the fluxes $\dot E^{H^+}$ and $\dot E^{H^-}$ do not exactly compensate for each other, leaving a net flux at the ECO radius that can be resonantly excited. Near the resonances and at high frequencies, this flux is comparable to $\dot E^\infty$ and contributes significantly to the GW phase.
\begin{figure*}[ht]
\centering
\begin{tabular}{cc}
 \includegraphics[width=0.42\textwidth]{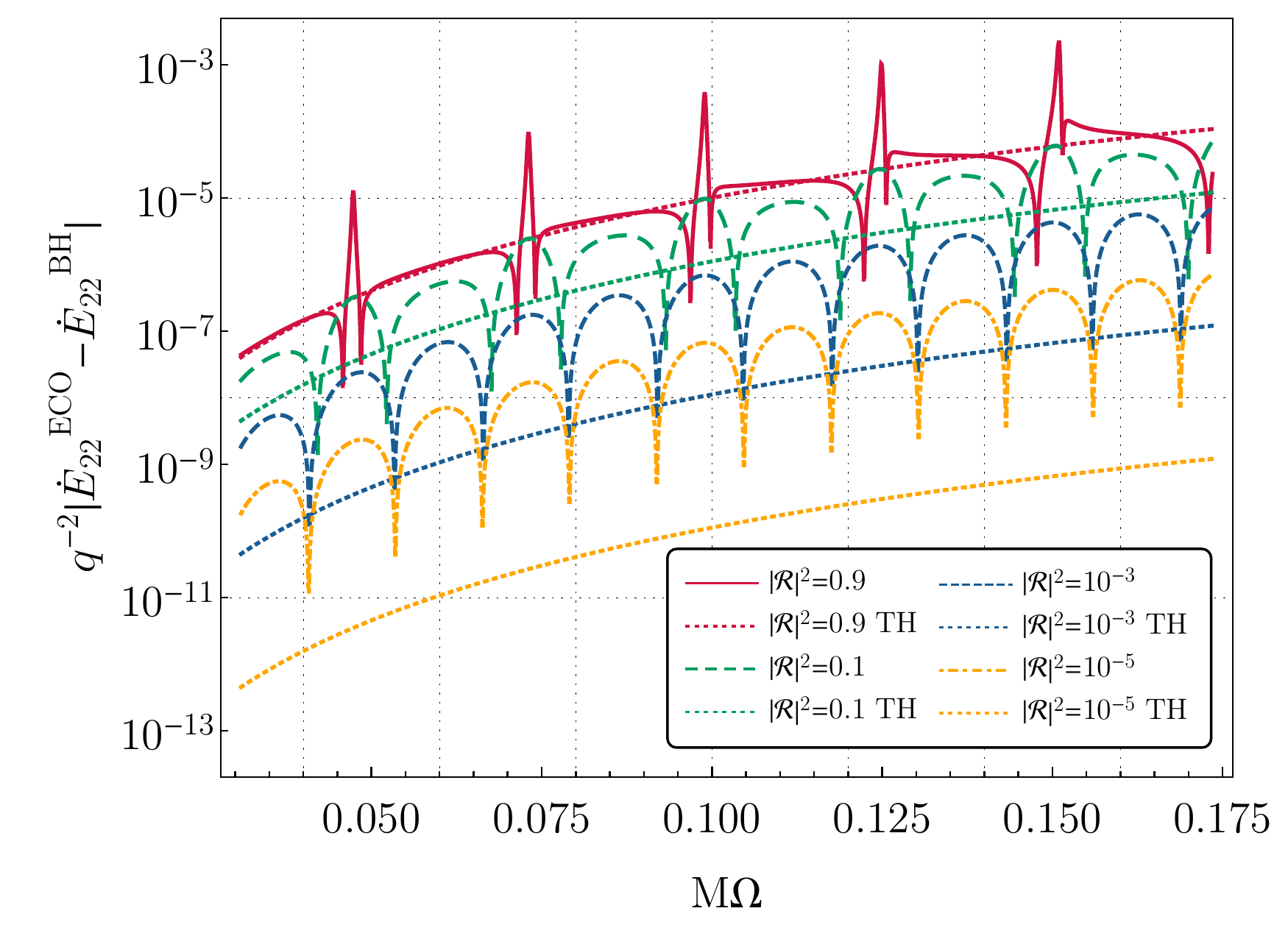} &
 \includegraphics[width=0.24\textwidth]{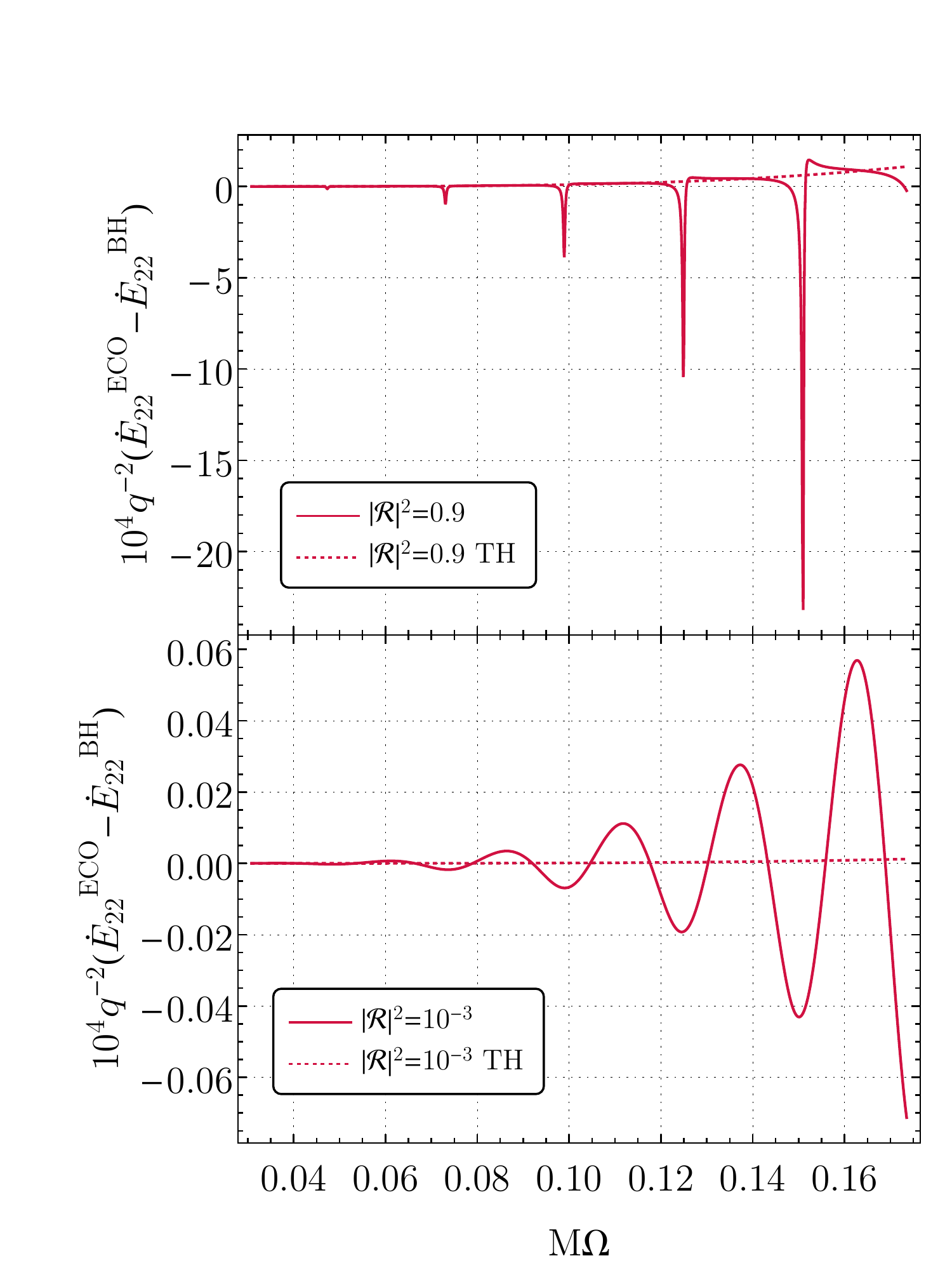}
\includegraphics[width=0.24\textwidth]{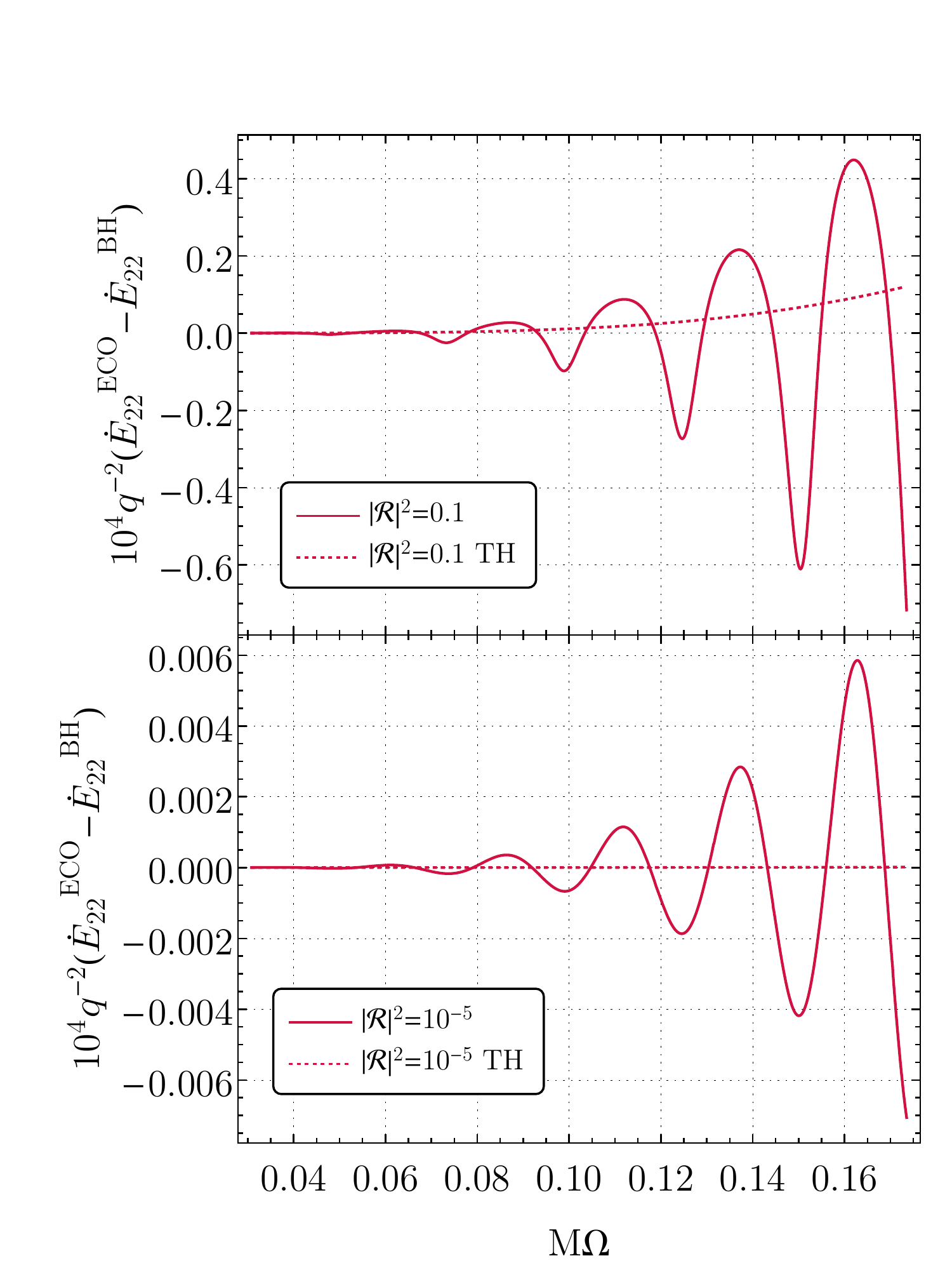}
\end{tabular}
\caption{
Difference between the total energy flux of the $\ell=m=2$ mode in the ECO case with respect to the BH case.
Left panel: absolute value of the difference for $\chi=0.8$, $\epsilon=10^{-10}$, and several values of the reflectivity. The dotted lines are the estimated differences in the total energy flux due to the absence of tidal heating relative to the BH case as described in Ref.~\cite{Datta:2019epe}.
Right panels: same as the left panel but without the absolute value and in a linear scale, to appreciate the change of sign during the oscillations associated with the resonances.
} 
\label{fig:difference}
\end{figure*}

Finally, in Fig.~\ref{fig:flux2} we also show the fluxes at infinity and at the ECO radius computed with the simplified model of Ref.~\cite{Datta:2019epe}, i.e., by artificially removing a fraction $(|{\cal R}|^2)$ of the tidal heating from a standard EMRI flux around a central Kerr BH. We observe that the energy flux at infinity in this case is similar to the exact result, except for the presence of the resonances (which are absent in the model of Ref.~\cite{Datta:2019epe}). On the other hand, the energy flux at the radius can change significantly. Owing to the presence of the resonances, $\dot{E}^{\rm int}$ computed in Ref.~\cite{Datta:2019epe} is, roughly speaking, a sort of averaged value of the exact result. The latter is modulated by the presence of resonances, which can be as high as the flux at infinity.

In Fig.~\ref{fig:difference}, we show the differences between the total energy flux of the $\ell=m=2$ mode in the horizonless case with respect to the BH case. In particular, the left panel shows the absolute value of this quantity on a logarithmic scale in order to appreciate the relatively small numbers involved. In the right panel grid, we instead show the same quantity on a linear scale and without the absolute value to appreciate the change of sign during the oscillations.

For $|\mathcal{R}|^2 \approx 1$ the differences between the consistent model and the model of Ref.~\cite{Datta:2019epe} 
are due to two factors: the excitation of resonances and the (subleading) fact that the flux computation in the consistent model is  more accurate since it accounts for the fraction of the GWs that are reflected by the object and make their way to infinity rather than being reabsorbed by the particle, as implicitly assumed in \cite{Datta:2019epe}. 
Furthermore, for smaller values of the reflectivity, the difference between our consistent model and the simplified one is even more important. In this case, the resonances are suppressed in amplitude but still appear in the total energy flux with a larger width, as shown in the left panel of Fig.~\ref{fig:difference}. The right panel grid in Fig.~\ref{fig:difference} shows the oscillatory trend of the total energy flux in the horizonless case compared to the energy flux in Ref.~\cite{Datta:2019epe} for small reflectivities. The amplitude of the oscillations increases with the orbital angular frequency and decreases with the reflectivity. These oscillations are related to the resonances and, as we shall see in Sec.~\ref{sec:dephasing}, they can also contribute significantly to the GW phase for small values of ${\cal R}$.

Interestingly, when the superradiance condition, $\Omega<\Omega_H$, is met, the flux at the radius can be negative due to superradiant energy and angular-momentum extraction from the central object~\cite{Brito:2015oca}. Since $\dot{E}^{\rm int}$ and $\dot E^\infty$ have opposite signs, it will be interesting to determine whether they can exactly compensate for each other at some given frequency, giving rise to a total zero flux and hence to ``floating'' orbits~\cite{Kapadia:2013kf,Cardoso:2011xi}. As clear from Fig.~\ref{fig:flux2}, limited to the case of the $\ell=m=2$ mode only such orbits would exist. However, they exist only near high-frequency resonances, where $\dot{E}^{\rm int}$ (which is typically subdominant) can be as large (in absolute value) as $\dot E^\infty$. When including the contribution of $\ell>2$ multipoles, we find that the total flux at infinity is larger than the flux at the radius because modes with different $(\ell,m)$ are resonantly excited at different frequencies. The net result is that the total flux, $\dot E^\infty+ \dot{E}^{\rm int}$, is positive overall and the orbit shrinks during the adiabatic evolution.

\subsection{Dephasing}\label{sec:dephasing}
%
\begin{figure}[ht]
\centering
\includegraphics[width=0.49\textwidth]{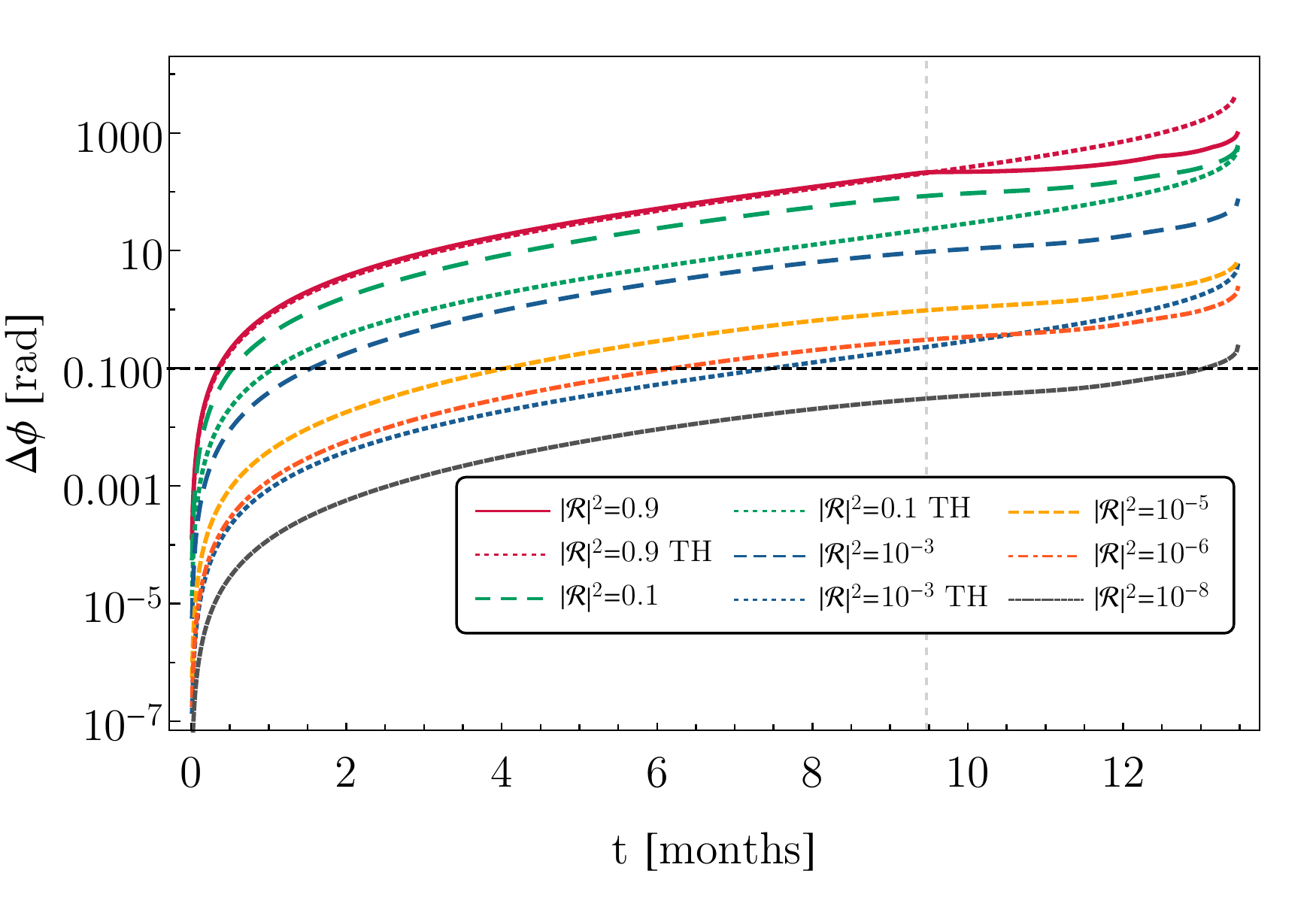}
\caption{GW dephasing between the BH and the ECO case as a  function of time for $\chi=0.8$, $q=3 \times 10^{-5}$, $\epsilon=10^{-10}$, and several values of reflectivity. The dotted lines show the dephasing due to the absence of tidal heating relative to the BH case as in Ref.~\cite{Datta:2019epe}.
The vertical dashed line denotes the time corresponding to a resonant orbital frequency. The horizontal line is a reference value $\Delta \phi=0.1\,{\rm rad}$~\cite{Lindblom:2008cm,Bonga:2019ycj}.
} 
\label{fig:dephasing}
\end{figure}

With the total flux at hand until $\ell_{\rm max}=12$,
we now study the dephasing between a horizonless compact object and the standard Kerr case. 
This is shown in Fig.~\ref{fig:dephasing} for a fiducial binary with $M=10^6 M_\odot$, $\mu=30M_\odot$, $\chi=0.8$, and $\epsilon=10^{-10}$. 
We analyze different values of the reflectivity $|{\cal R}|^2$ and for each of them we compare our exact result to that of the model adopted in Ref.~\cite{Datta:2019epe}.
As expected, the dephasing increases monotonically in time (except possibly when a resonance is crossed, in which case the dephasing can have an antispike and decreases near the resonant frequency; see Appendix~\ref{app:dephasing_compactness}), and also as a function of the reflectivity.
When $|{\cal R}|^2\approx 1$, the difference with respect to the model adopted in Ref.~\cite{Datta:2019epe} is small 
until the inspiral moves across a resonance. In particular, for $|\mathcal{R}|^2=0.9$, the dephasing with respect to our exact results deviates from the dephasing due to the absence of tidal heating at $t=9.47\,{\rm months}$ (marked in Fig.~\ref{fig:dephasing} as a dashed vertical line) due to the presence of a $\ell=m=2$ resonance at $M\Omega=0.0473$ with $M\omega_I = -4.22\times 10^{-5}$. Subsequent resonances are excited at later times.

The simplified model of Ref.~\cite{Datta:2019epe} and the exact result differ significantly for small reflectivities even if the resonances are less evident.
This is due to several factors: the energy fluxes at the ECO radius and at infinity display some differences in the two models since a fraction of energy is reflected by the object and leaves the system; moreover, both fluxes (at the radius and at infinity) can be resonantly excited only in our consistent model, and these resonances contribute significantly to the GW phase for intermediate values of ${\cal R}$.
The dephasing in the consistent model is always larger than the dephasing with tidal heating only (for very small values of ${\cal R}$, the two models differ, but both produce a tiny dephasing, as expected).
The dephasing depends mildly on the compactness of the object; see Appendix~\ref{app:dephasing_compactness} for an analysis of this contribution.

\subsection{Overlap}\label{sec:resoverlap}

%
\begin{figure}[ht]
\centering
\includegraphics[width=0.49\textwidth]{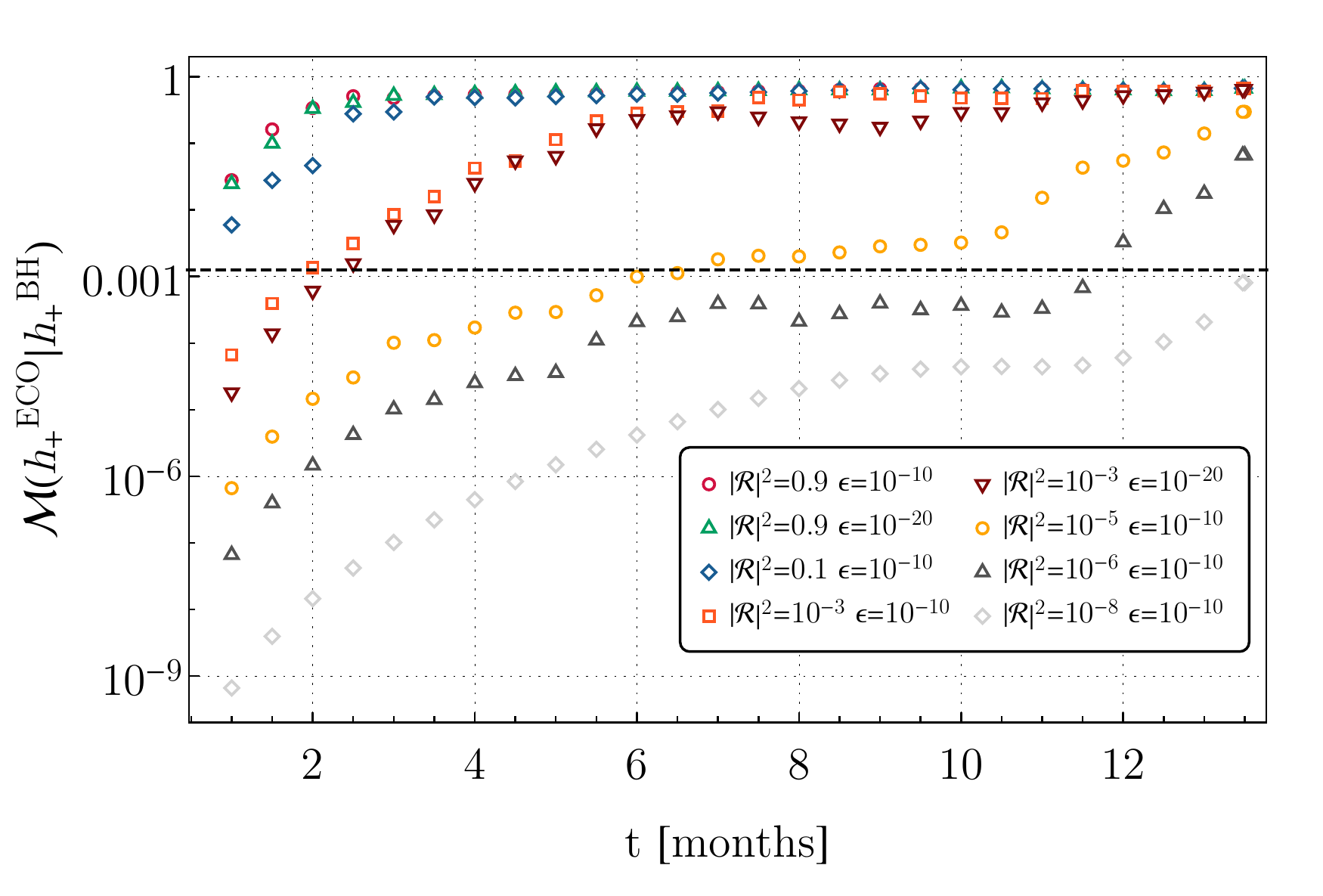}
\caption{Mismatch between the plus polarization of the waveforms with a central ECO and a central BH as a function of time, for $\chi=0.8$, $q=3 \times 10^{-5}$, and several values of the reflectivity.} 
\label{fig:overlap}
\end{figure}

In Fig.~\ref{fig:overlap} we show the mismatch $\mathcal{M}\equiv 1-{\cal O}$ between the waveforms in the ECO case and in the Kerr case with the same mass and spin for various values of ${\cal R}$ and two choices of $\epsilon$.

As is clear from the plot, the value of $\epsilon$ does not affect the mismatch significantly as long as $\epsilon\ll1$. As with the dephasing presented above, the mismatch is larger for the consistent model, especially at small reflectivity, as can be appreciated by comparing Fig.~\ref{fig:overlap} to the corresponding plot in Ref.~\cite{Datta:2019epe}.
As a useful rule of thumb, two waveforms can be considered indistinguishable 
for parameter-estimation purposes if $\mathcal{M}\lesssim 
1/(2\rho^2)$, where $\rho$ is the signal-to-noise ratio of the true signal~\cite{Flanagan:1997kp,Lindblom:2008cm}. For an EMRI with $\rho\approx 20$ ($\rho\approx 100$) one has $\mathcal{M}\lesssim 10^{-3}$ ($\mathcal{M}\lesssim 
5\times 10^{-5}$). In Fig.~\ref{fig:overlap} the more conservative threshold $\mathcal{M}= 10^{-3}$ is denoted with a 
dashed horizontal line. Exceeding this threshold is a necessary but not  sufficient condition  for a deviation to be detectable.
This level of mismatch is quickly exceeded after less than one year of data, even for small values of the reflectivity. For example, for the fiducial case considered in Fig.~\ref{fig:overlap} ($\chi=0.8$, $M=10^6M_\odot$, and $\mu=30 M_\odot$), and assuming that $\rho=20$, the threshold is exceeded after roughly one year unless
\begin{equation}
 |{\cal R}|^2 \lesssim 10^{-8}\,. \label{bound}
\end{equation}
Note, however, that the above bound is based solely on the mismatch calculation and does not take into account, e.g., correlations with other waveform parameters. Rigorous parameter estimation is necessary to derive an accurate projected upper bound (in the case of no detection). This interesting analysis goes beyond our scope and is left for future work.

\subsection{A case study: EMRI constraints on Boltzmann reflectivity}\label{sec:Boltzmann}

%
\begin{figure*}[ht]
\centering
\includegraphics[width=0.49\textwidth]{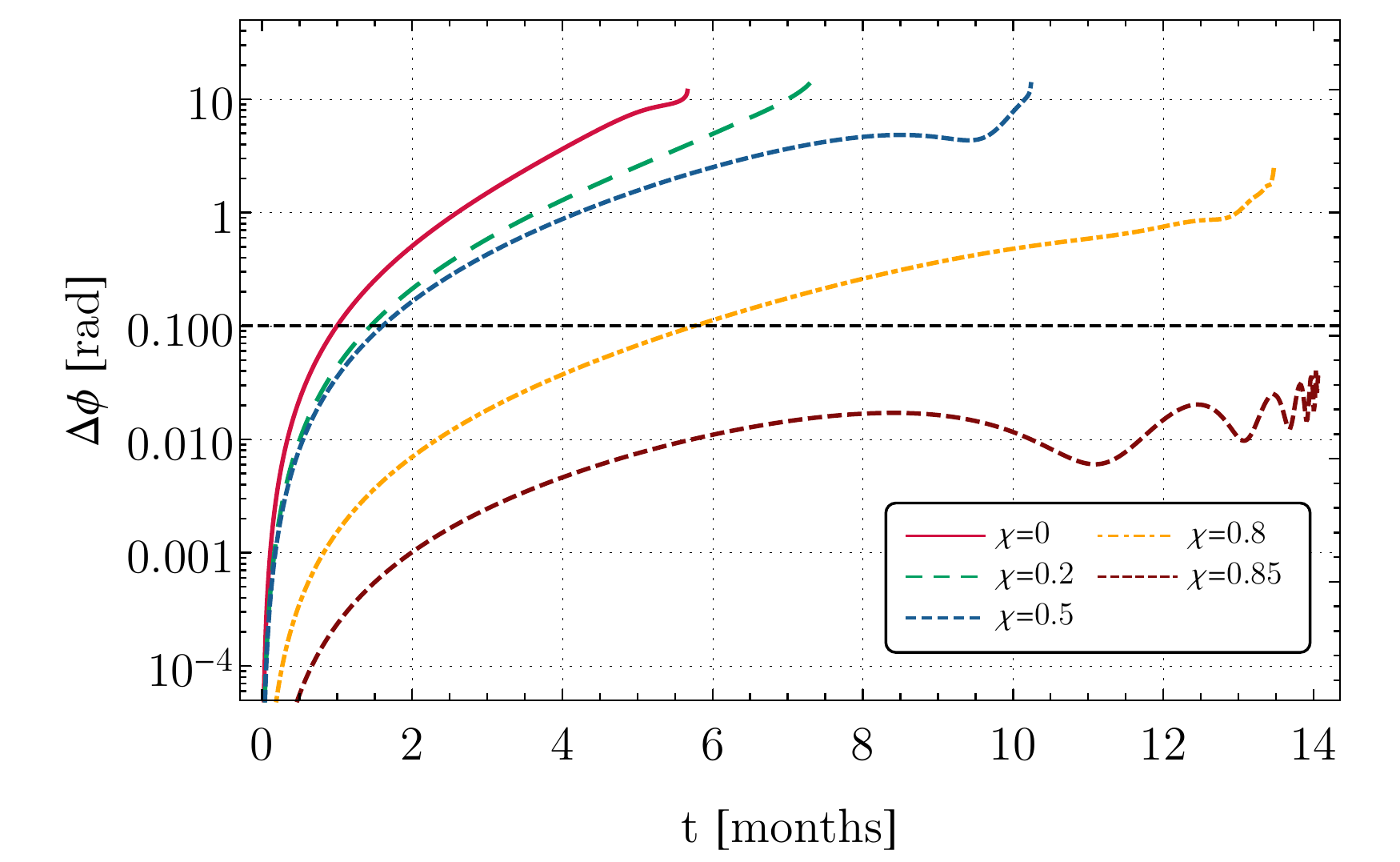}
\includegraphics[width=0.49\textwidth]{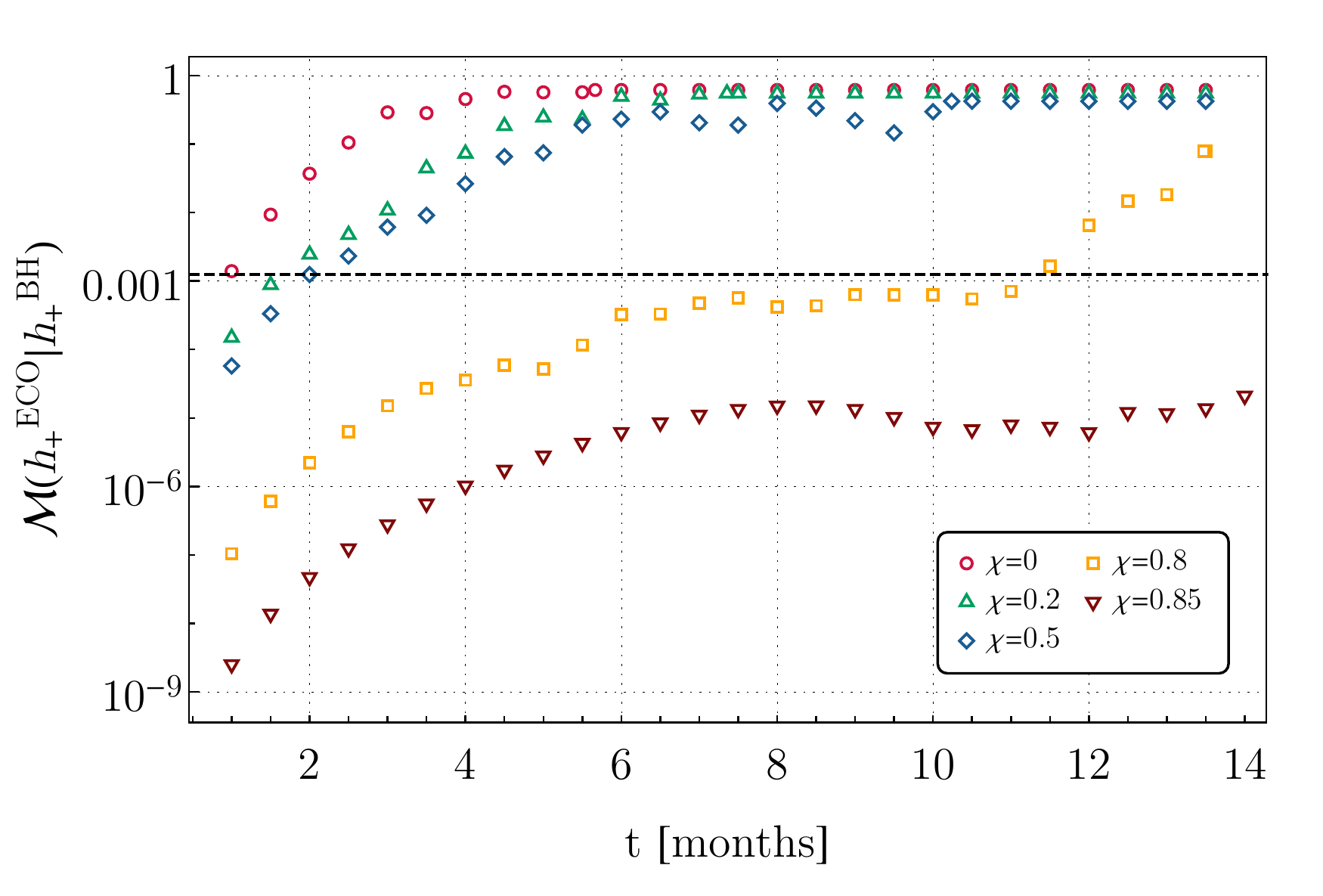}
\caption{Left panel: GW dephasing between the Kerr case and a quantum BH horizon with Boltzmann reflectivity [in Eq.~\eqref{Boltzmann}], $\epsilon=10^{-10}$, $q=3 \times 10^{-5}$, and various values of the spin as a function of time. Right panel: mismatch between the plus polarization of the waveform with a central quantum BH horizon with Boltzmann reflectivity and a central BH as a function of time for several values of the primary spin.}
\label{fig:Boltzmann}
\end{figure*}

Although thus far we have considered only the case in which $|{\cal R}|^2={\rm const}$, an advantage of our framework is that the reflectivity coefficient can be a generic complex function of the model's parameters and the frequency. We now consider a specific model for the ECO reflectivity. In particular, we shall assume a model recently proposed to describe quantum BH horizons that gives rise to ``Boltzmann'' reflectivity~\cite{Oshita:2019sat,Wang:2019rcf},
\begin{equation}
 \mathcal{R}(\omega) = e^{-\frac{|k|}{2 T_H}} \,, \label{Boltzmann}
\end{equation}
where $T_H = \frac{r_+ - r_-}{4 \pi (r_+^2 + a^2)}$ is the Hawking temperature of a Kerr BH. In this model, the reflectivity depends explicitly on the spin and the frequency. Furthermore, it provides sufficient absorption to quench the ergoregion instability~\cite{Oshita:2019sat}.
Note that Eq.~\eqref{Boltzmann} can also contain a phase term that depends on the specific model and the perturbation function on which the corresponding boundary condition is imposed~\cite{Oshita:2019sat,Wang:2019rcf,Xin:2021zir}. For simplicity, here we shall neglect such a phase term, which does not affect our analysis anyway.~\footnote{Recently, Refs.~\cite{Chen:2020htz,Xin:2021zir} proposed an alternative model for ECO reflectivity that is related to the tidal response of the ECO to external curvature perturbations. In this model, the reflectivity contains extra terms that multiply the Boltzmann factor.}

Figure~\ref{fig:Boltzmann} shows the dephasing (left panel) and the overlap (right panel) obtained in the Boltzmann reflectivity model compared to the classical BH case.
An interesting feature of this model is that there is no free parameter that continuously connects it to the classical Kerr case, so there is a concrete chance to rule it out with observations or to provide evidence for it. 
Interestingly, owing to its spin dependence, the Boltzmann reflectivity is much smaller at the relevant orbital frequencies when the central object is highly spinning. 
 Therefore, as shown in Fig.~\ref{fig:Boltzmann},
 the dephasing and the mismatch with respect to the standard Kerr BH case are very small when $\chi\gtrsim0.8$. The oscillatory trend in the dephasing is due to the contribution of high-frequency resonances appearing at late times.

\section{Conclusion} \label{sec:discussion}
EMRIs will be unparalleled probes of fundamental physics and unique sources for the LISA mission, and evolved concepts thereof~\cite{Baibhav:2019rsa}.
Developing a consistent model of a partially absorbing ECO, we studied the signal emitted by a point particle in a circular motion.
The EMRI dynamics are affected by the modified boundary conditions at the object's surface, which give rise to modified tidal heating, modified fluxes, and resonant QNM excitations in a consistent fashion.
We showed that the GW emission and orbital dynamics in the consistent model are quite rich: in addition to some quantitative differences with respect to the simplified model studied in Ref.~\cite{Datta:2019epe}, there are also qualitatively new features such as resonances that might give a relevant contribution to the GW phase in some regions of the parameter space. In principle, these resonances could also jeopardize detection if not suitably accounted for in the waveform. 

Overall, we found that the already very stringent potential bounds derived in Ref.~\cite{Datta:2019epe} can be further improved by some orders of magnitude by taking into account a consistent ECO model. These projected constraints suggest that EMRI could place the strongest bounds on the reflectivity of supermassive objects, orders of magnitude more stringent than those potentially coming from echo searches in the postmerger phase of comparable mass coalescences~\cite{Testa:2018bzd,Maggio:2019zyv}.
In particular, we showed that an EMRI detection is potentially sensitive to an effective reflectivity of the central supermassive object as small as
$|{\cal R}|^2\sim {\cal O}(10^{-8})$. As a reference, we remind the reader that in the BH case the reflectivity is exactly zero and that for a neutron star it is practically unity,\footnote{
Note that this reflectivity refers to the reduced one-dimensional scattering problem, where $|{\cal R}|^2=1$ simply corresponds to the fact that ``everything that goes in eventually comes out''. This is the case when the absorption of the incoming radiation is negligible since in that case (from a three-dimensional perspective) radiation simply passes unperturbed across the object. In terms of the object's viscosity, the reflectivity in the high-frequency or $\epsilon\to0$ limit reads~\cite{Abedi:2020ujo,Maggio:2020jml}
\begin{equation}
 |{\cal R}|^2=\frac{(\eta - \eta_{\rm BH})^2}{(\eta + \eta_{\rm BH})^2}=1-{\cal O}(\eta/\eta_{\rm BH})\,,
\end{equation}
where $\eta_{\rm BH} = 1/(16\pi)$ is the effective viscosity of a BH within the BH membrane paradigm. Hence, if $\eta\ll\eta_{\rm BH}$ (an excellent approximation for GWs interacting with ordinary matter~\cite{1971ApJ...165..165E}), the effective reflectivity is practically unity, i.e., GWs are not absorbed and tidal heating is practically zero. 
} even when accounting for dissipation due to viscosity~\cite{1971ApJ...165..165E,Maggio:2017ivp,Maggio:2018ivz}.
Furthermore, we showed that this unique sensitivity to small reflectivity coefficients can be used to constrain specific ECO models, such as those of quantum BH horizons
featuring Boltzmann reflectivity~\cite{Oshita:2019sat,Wang:2019rcf}.
Our approach is general and the reflectivity coefficient can be an arbitrarily complex function of the model parameters and the frequency, so the same analysis can be applied to other specific ECO models; see, e.g.~\cite{Chen:2020htz,Xin:2021zir}.

However, the above conclusion is based on several simplifications that should be relaxed in future work. In particular, we focused on circular, equatorial orbits, while EMRIs are expected to be eccentric and nonplanar, introducing two further parameters (the eccentricity and the Carter constant) to the description of the inspiral. Future work should also include leading-order self-force effects~\cite{Barack:2009ux,Poisson:2011nh,Pound:2021qin}, which are needed for accurate parameter estimation with EMRIs~\cite{Hinderer:2008dm}. Both can be done with minor adjustments to the code of Refs.~\cite{vandeMeent:2014raa, vandeMeent:2015lxa, vandeMeent:2016pee, vandeMeent:2017bcc} described here. Finally, the upper bounds estimated here are based on the overlap calculation, and therefore neglect possible correlations among the waveform parameters, which is particularly relevant for generic orbits and a  relatively small signal-to-noise ratio.
From the parameter-estimation point of view, it is important to develop modified kludge waveforms to include ECO effects in a practical way or, more ambitiously, to perform accurate data analyses using exact waveforms (either using the Fisher-information matrix or, ideally, a Bayesian inference), extending recent work in the context of standard waveforms~\cite{Chua:2019wgs,Katz:2021yft,Piovano:2021iwv}.

{\bf Note added:} After this work was completed, we became aware of a paper by Norichika Sago and Takahiro Tanaka, who independently studied the same problem, obtaining very similar results~\cite{Sago:2021iku}.

\begin{acknowledgments}
The authors are grateful to Richard Brito, Sayak Datta, Simone Mastrogiovanni, and Gabriel Andres Piovano for useful discussions.
Numerical computations were performed at the Vera cluster of the Amaldi Research Center funded by the MIUR program ``Dipartimento di 
Eccellenza'' (CUP:~B81I18001170001).
E.~M. and P.~P. acknowledge the financial support provided under the
European Union's H2020 ERC, Starting Grant agreement no.~DarkGRA--757480. We also acknowledge support under the MIUR PRIN and FARE programmes (GW- NEXT, CUP: B84I20000100001).
\end{acknowledgments}

\appendix

\section{Dephasing in the nonspinning case}\label{app:dephasing_spin0}

\begin{figure}[ht]
\centering
\includegraphics[width=0.49\textwidth]{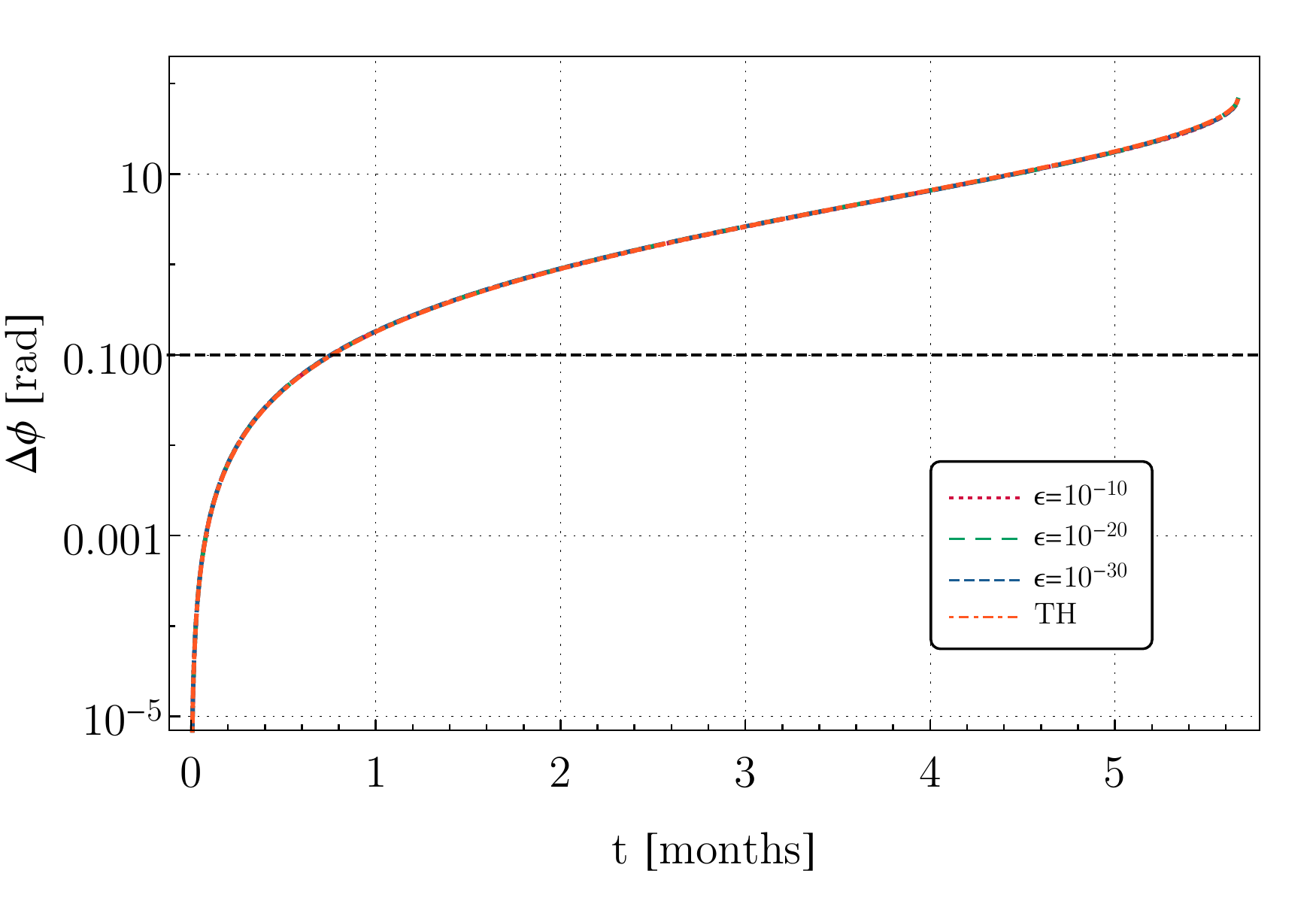}
\caption{Dephasing as a function of time in the case of a nonspinning, perfectly reflecting ECO relative to the Schwarzschild BH case for different values of the compactness parameters $\epsilon$ and $q=3 \times 10^{-5}$. In this case, the resonances in the flux at infinity do not contribute to the dephasing, which is well approximated by the simplified model of Ref.~\cite{Datta:2019epe}.} 
\label{fig:dephasing_spin0}
\end{figure}

For completeness, here we show the dephasing in the case of a nonspinning, perfectly reflecting ECO relative to the Schwarzschild BH case for different values of the compactness parameters $\epsilon$. Figure~\ref{fig:dephasing_spin0} shows that the dephasing essentially does not depend on $\epsilon$ and is not affected by the resonances, which in the nonspinning case are too narrow to be efficiently excited.
%

\section{Fluxes and dephasing as function of the compactness}\label{app:dephasing_compactness}
%

\begin{figure}[ht]
\centering
\includegraphics[width=0.49\textwidth]{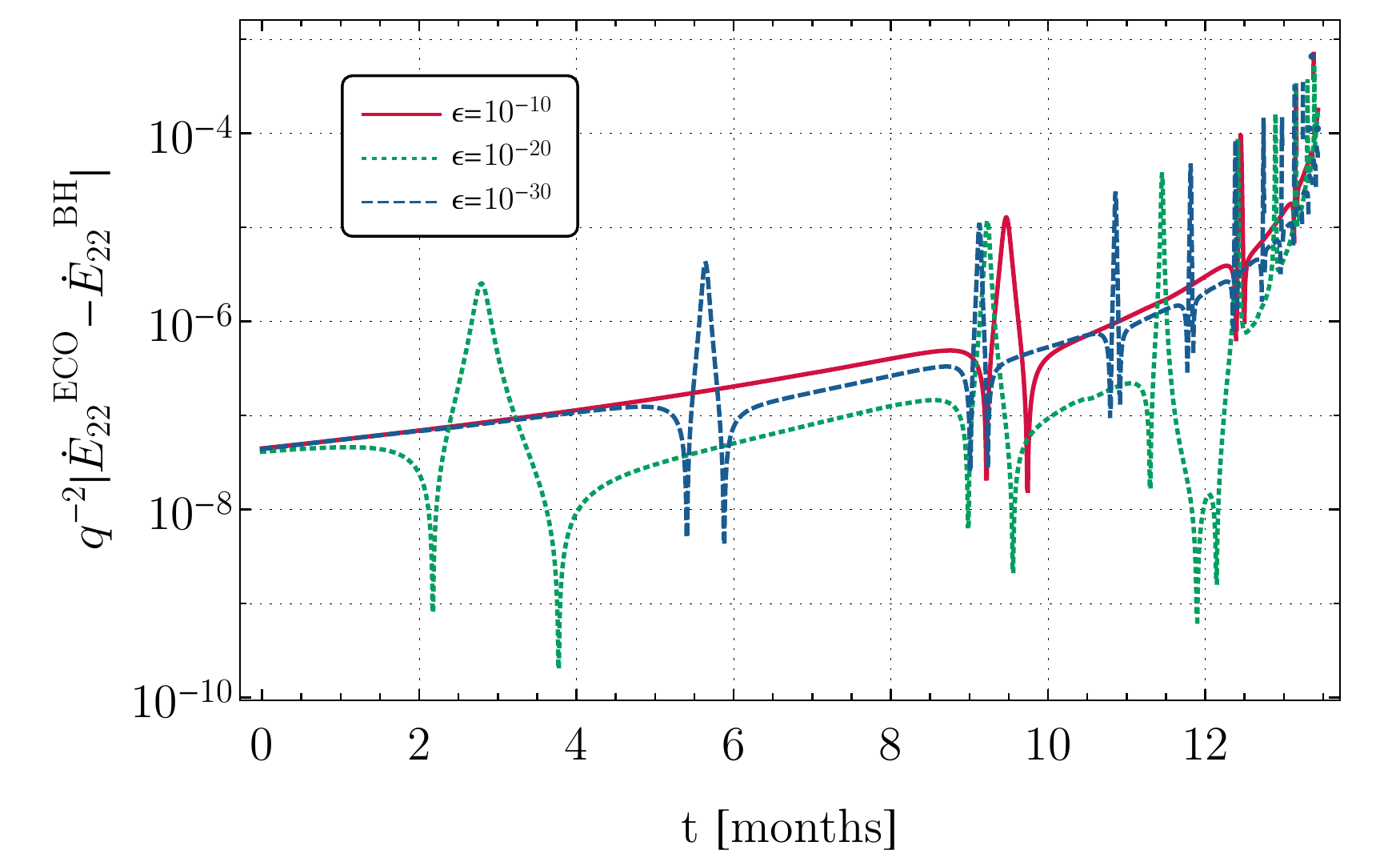}
\caption{Resonances in the $\ell=m=2$ energy flux for an ECO with $\chi=0.8$, $|\mathcal{R}|^2=0.9$, and several values of $\epsilon$ as a function of time.}
\label{fig:resonances_time}
\end{figure}

In Fig.~\ref{fig:resonances_time}, we show the difference between the ECO and Kerr BH total energy fluxes for several values of $\epsilon$ as a function of time. We note that, as $\epsilon$ decreases, more resonances appear, and they also appear at lower frequencies. The first low-frequency resonances might give a large contribution to the phase since the orbital evolution is slower at low frequency and the particle can spend more time moving across the resonance. On the other hand, in our ECO model the width of each resonance is proportional to $\omega_I \sim \omega_R^{2\ell+2}$, and therefore low-frequency resonances are also more narrow. The two effects are competitive and the actual contribution of a resonance on the GW phase depends on the specific parameters of the configuration.

\begin{figure}[ht]
\centering
\includegraphics[width=0.49\textwidth]{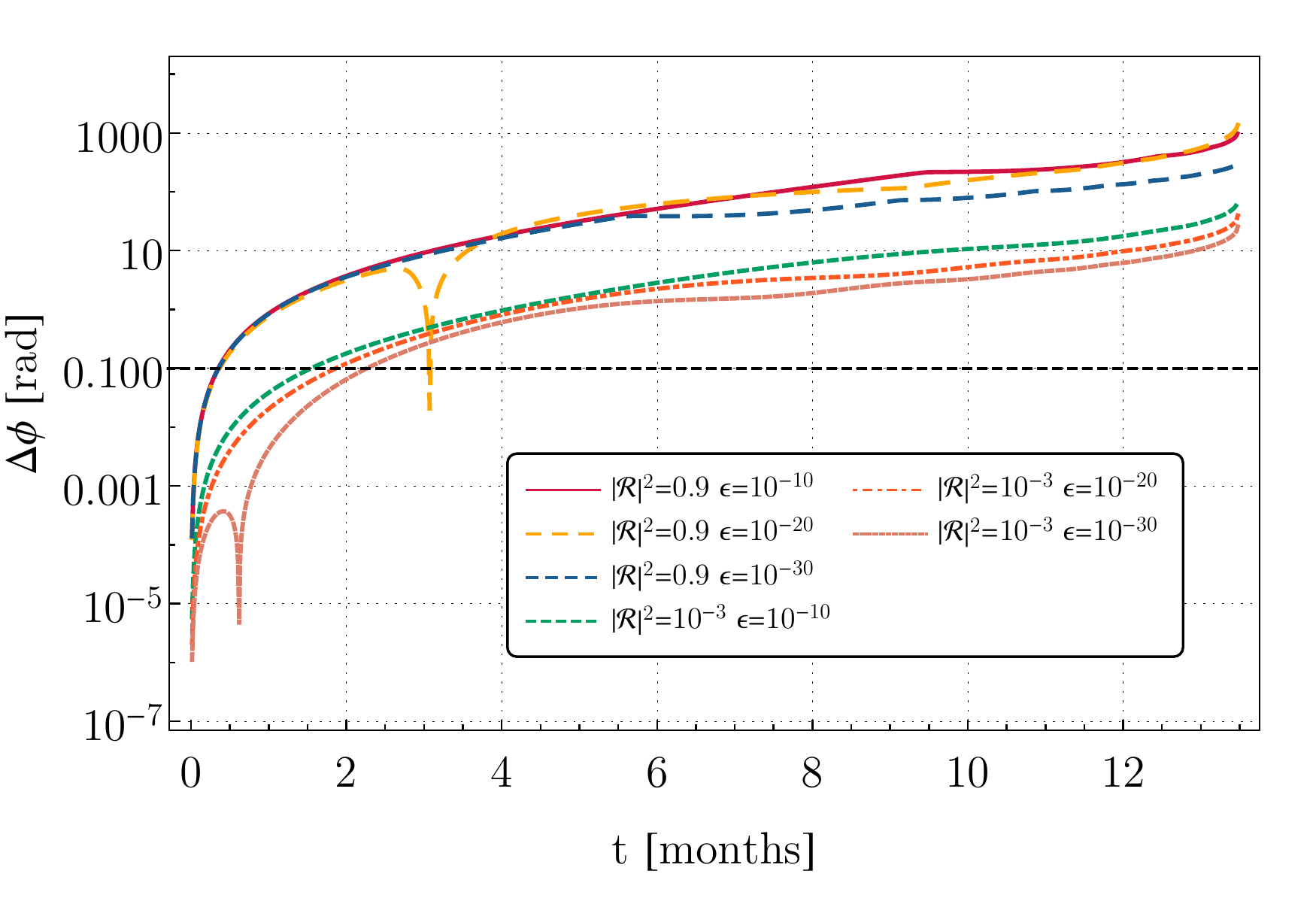}
\caption{GW dephasing between the BH and the ECO case as a  function of time for $\chi=0.8$, $q=3 \times 10^{-5}$, and several values of $\epsilon$.}
\label{fig:dephasing_compactness}
\end{figure}

Finally, in Fig.~\ref{fig:dephasing_compactness} we show the dephasing for some values of $|{\cal R}|^2$ and $\epsilon$. The dependence on $\epsilon$ is mild, except for the possible excitation of the resonances, whose impact depends on the specific values of $\chi$, $\epsilon$, and ${\cal R}$.

\bibliographystyle{apsrev4}
\bibliography{References}

\end{document}